\documentclass[lettersize,journal]{IEEEtran}
\usepackage{amsmath,amsfonts,amssymb}
\usepackage{algorithm}
\usepackage{enumerate}
\usepackage{algpseudocode}
\usepackage{array}
\newcolumntype{?}{!{\vrule width 1pt}}

\newtheorem{remark}{\textbf{Remark}}
\usepackage[caption=false, labelformat=parens, subrefformat=parens]{subfig}
\usepackage{textcomp}
\usepackage{stfloats}
\usepackage{url}
\usepackage{diagbox}
\usepackage{verbatim}
\usepackage{graphicx}
\usepackage{cite}
\usepackage{booktabs}
\usepackage[hidelinks]{hyperref} 
\usepackage{makecell}
\usepackage{xcolor}
\usepackage[export]{adjustbox}
\usepackage{array, multirow}
\usepackage{tabularx}
\usepackage[switch]{lineno}
\newcommand{\multiline}[2]{%
  \begin{tabularx}{#1\textwidth}[t]{@{}X@{}}
    #2
  \end{tabularx}}
\begin{document}
% \linenumbers
\title{radarODE-MTL: A Multi-Task Learning Framework with Eccentric Gradient Alignment for Robust Radar-Based ECG Reconstruction}

\author{Yuanyuan~Zhang, Rui~Yang, \IEEEmembership{Senior~Member,~IEEE}, \\ Yutao~Yue, \IEEEmembership{Senior~Member,~IEEE}, Eng~Gee~Lim, \IEEEmembership{Senior~Member,~IEEE % <-this % stops a space
\thanks{This research has been approved by the University Ethics Committee of Xi'an Jiaotong-Liverpool University with proposal number ER-SAT-0010000090020220906151929, and is partially supported by Suzhou Science and Technology Programme (SYG202106), Jiangsu Industrial Technology Research Institute (JITRI) and Wuxi National Hi-Tech District (WND). \textit{(Corresponding authors: Rui Yang, Yutao Yue.)}}
\thanks{Yuanyuan Zhang is with the School of Advanced Technology, Xi'an Jiaotong-Liverpool University, Suzhou, 215123, China, the Department of Electrical Engineering and Electronics, University of Liverpool, Liverpool, L69 3GJ, United Kingdom, and also with the Institute of Deep Perception Technology, JITRI, Wuxi, 214000, China (email: Yuanyuan.Zhang16@student.xjtlu.edu.cn).}
\thanks{Rui Yang and Eng Gee Lim are with the School of Advanced Technology, Xi'an Jiaotong-Liverpool University, Suzhou, 215123, China (email: R.Yang@xjtlu.edu.cn; Enggee.Lim@xjtlu.edu.cn).}
\thanks{Yutao Yue is with the Thrust of Artificial Intelligence and Thrust of Intelligent Transportation, The Hong Kong University of Science and Technology (Guangzhou), Guangzhou 511400, China, and also with the Institute of Deep Perception Technology, JITRI, Wuxi 214000, China. (email: yutaoyue@hkust-gz.edu.cn).}}}% <-this % stops a space 

\maketitle

\begin{abstract}
Millimeter-wave radar is promising to provide robust and accurate vital sign monitoring in an unobtrusive manner. However, the radar signal might be distorted in propagation by ambient noise or random body movement, ruining the subtle cardiac activities and destroying the vital sign recovery. In particular, the recovery of electrocardiogram (ECG) signal heavily relies on the deep-learning model and is sensitive to noise. Therefore, this work creatively deconstructs the radar-based ECG recovery into three individual tasks and proposes a multi-task learning (MTL) framework, radarODE-MTL, to increase the robustness against consistent and abrupt noises. In addition, to alleviate the potential conflicts in optimizing individual tasks, a novel multi-task optimization strategy, eccentric gradient alignment (EGA), is proposed to dynamically trim the task-specific gradients based on task difficulties in orthogonal space. The proposed radarODE-MTL with EGA is evaluated on the public dataset with prominent improvements in accuracy, and the performance remains consistent under noises. The experimental results indicate that radarODE-MTL could reconstruct accurate ECG signals robustly from radar signals and imply the application prospect in real-life situations. The code is available at: http://github.com/ZYY0844/radarODE-MTL.
\end{abstract}


\section{Introduction}
Electrocardiogram (ECG) signal is commonly recognized as the golden standard in cardiac monitoring compared with other vital signs (e.g., heart rate, photoplethysmography), because ECG describes the fine-grained cardiac activities, such as atrial/ventricular depolarization/repolarization, through the featured waveform (i.e., PQRST peaks) and is crucial to the diagnosis of cardiovascular diseases~\cite{swift2021stop}. The traditional ECG measurement relies on the adhesive electrode patches with wired connections to the monitor to provide real-time and accurate ECG signals and is mainly used in clinical scenarios due to the cumbersome apparatus. However, the contact-based ECG collection is unfriendly to long-term monitoring and is not applicable to daily wellness monitoring~\cite{chen2022contactless}. Recently, radar has become a promising contactless sensor to provide non-invasive and accurate ECG monitoring by using advanced signal-processing algorithm and deep neural network~\cite{chen2022contactless,wu2023contactless,wang2023ecg,zhang2024radarODE}.

The trials on the radar-based ECG recovery can be categorized into two paradigms. The first paradigm only performs the extraction of high-resolution cardiac mechanical activities to produce quasi-ECG signals, omitting the morphological ECG features while maintaining certain fine-grained features. For example, the mostly adopted quasi-ECG signal only preserves R and T peaks and can be realized by signal decomposition~\cite{dong2024robust} or state estimation~\cite{ji2022rbhhm,xia2021radar}. In contrast, the second paradigm aims to reconstruct the ECG waveform as measured by clinical apparatus, because the doctor and ECG analysis toolbox all rely on the shape of ECG to make diagnosis~\cite{makowski2021neurokit2}. However, decoupling the ECG signal from the measured radar signal requires establishing an extremely complex model from the perspective of electrophysiology (i.e., excitation-contraction coupling~\cite{swift2021stop}), and the existing research can only leverage deep learning methods to learning such domain transformation from the dataset containing numerous radar/ECG pairs~\cite{chen2022contactless,wu2023contactless,wang2023ecg,zhang2024radarODE}.

In the literature, radar-based ECG waveform recovery has been achieved based on various deep-learning architectures, such as convolutional neural network (CNN)~\cite{chen2022contactless,zhang2024radarODE}, long short-term memory (LSTM) network~\cite{ji2022rbhhm}, and Transformer~\cite{chen2022contactless,wu2023contactless}. However, the noise robustness of the deep-learning framework is rarely investigated in the literature, especially for the random body movement (RBM) noise that is inevitable in contactless monitoring and has orders of magnitude larger than cardiac activities. The existing work either discarded the data during the RBM~\cite{wang2023ecg} or reported the heavy distortion as the future work~\cite{chen2022contactless}. Additionally, the existing deep-learning methods are also blamed for being purely data-drove as a black box and the transformation between cardiac mechanical and electrical activities lacks the theoretical explanation~\cite{zhang2024radarODE}. 

Based on the limitations of the existing methods, it is necessary to provide a feasible model that explains the transformation inside radar-based long-term ECG recovery and is also robust to real-life noises. Therefore, this work proposes to deconstruct the radar-based ECG reconstruction into three individual tasks as a multi-task learning (MTL) problem to extract cardiac features with different levels of granularity, i.e., coarse features: heartbeat detection and cardiac cycle timing; fine-grained feature: ECG waveform. However, another consequent problem is to simultaneously optimize three individual tasks under the MTL paradigm, because the optimization of one task may degrade the performance of the others~\cite{lin2023libmtl,guan2024talk2radar}. 

In the literature, MTL is a widely-used deep learning paradigm in various fields such as scene understanding~\cite{silberman2012indoor,yeshwanth2023scannet++}, autonomous driving~\cite{yao2023waterscenes} and speech/text processing~\cite{liu2024primary}. However, the MTL paradigm has never been applied in radar-based ECG recovery, and the existing MTL optimization strategies cannot fairly optimize all the tasks due to the imbalanced task difficulties~\cite{guo2018dynamic}. In this work, the difficulty of extracting the ECG waveform is much higher than the other two, and simply applying the existing optimization strategies cannot achieve an ideal result with fair improvements on all tasks according to our initial experiments. 

Inspired by the above discussion, the contributions of this work can be concluded as:
\begin{itemize}
\item A novel optimization strategy called eccentric gradient alignment (EGA) is proposed for updating shared parameters in the MTL neural network, aiming to balance the intrinsic difficulty across tasks during network training and also prevent the negative transfer phenomenon. 
\item To the best of our knowledge, this is the first work that investigates the noise robustness in radar-based ECG recovery against constant or abrupt noise by modeling the cardiac domain transformation as three tasks. An end-to-end MTL framework named radarODE-MTL is accordingly proposed to realize these tasks and leverage adjacent cardiac cycles to compensate for the distorted one.
\item Sufficient experiments show that the proposed radarODE-MTL with EGA optimization strategy outperforms other frameworks and optimization strategies under various noise conditions and datasets, and the deconstructed tasks in radarODE-MTL could further improve the interpretability in radar-based ECG recovery.
\end{itemize}

The rest of the paper is organized as follows. Section~\ref{sec:bg} provides the background for radar-based ECG recovery and MTL optimization. The proposed radarODE-MTL framework with EGA strategy is elaborated in Section~\ref{sec:method}, and the experimental settings and results are shown in Section~\ref{sec:results}. At last, Section~\ref{sec:conclusions} concludes this paper with future work.

\section{Background and Problem Statement}\label{sec:bg}
This section will provide compact explanations of the domain transformation in ECG recovery and the optimization problem in MTL network, with the corresponding problem statements.

\subsection{Model for Domain Transformation and Problem Statement}
\subsubsection{Signal Model for Cardiac Mechanical Activities}
\begin{figure}[tb] 
    \centering 
    \includegraphics[width=0.8\columnwidth]{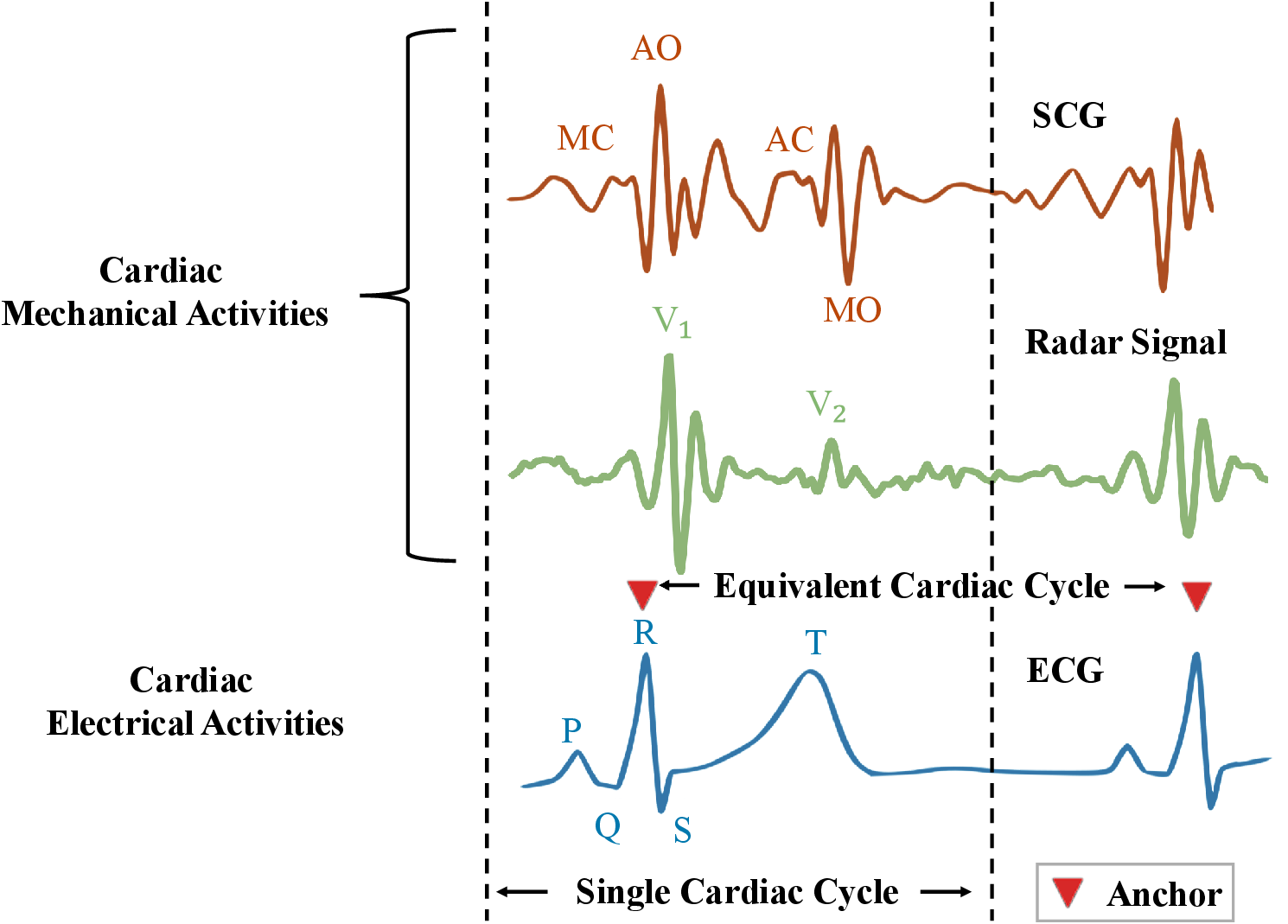}
    \caption{Relationships between cardiac mechanical and electrical activities, with single cardiac cycle and ECG anchors labeled.} 
    \label{fig:scg_ecg} 
\end{figure}

In radar-based ECG recovery, the baseband signal is normally pre-processed using bandpass filter, differentiator and digital beamforming to remove the background and respiration noise to enhance cardiac-related features~\cite{dong2024robust,chen2022contactless,shen2018respiration}. According to our previous work~\cite{zhang2024radarODE}, the fine-grained cardiac mechanical activities include aortic valve opening/closure (AO/AC) and mitral valve opening/closure (MO/MC), revealed by the corresponding prominent vibrations $v_1$ and $v_2$ as measured in radar signal $x(t)$ as depicted in Figure~\ref{fig:scg_ecg}. Therefore, the resultant radar signal $x(t)$ can be expressed for $K$ cardiac cycles as:

\begin{equation}\label{equ:vib_long}
x(t) = \sum_{k=1}^{K} v^k_1(t) + \sum_{k=1}^{K} v^k_2(t) + n_{abr}(t) + n_{con}(t)
\end{equation}
with
\begin{equation}\label{equ:t1}
\begin{aligned}
 v^k_1 (t) &= \mathrm{a}^k_1 \mathrm{cos}(2\pi f^k_1 t)\exp \left(-\frac{(t-T^k_1)^2}{{b^k_1}^2}\right)\\
v^k_2 (t) &= \mathrm{a}^k_2 \mathrm{cos}(2\pi f^k_2 t) \exp \left(-\frac{(t-T^k_2)^2}{{b^k_1}^2}\right)
 \end{aligned}
\end{equation}
where $a_1^k$, $b_1^k$ and $a_2^k$, $b_2^k$ jointly determine the amplitudes and lengths of the first and second prominent vibrations for $k^{th}$ cardiac cycle, $f_1^k$, $f_2^k$ are the corresponding central frequencies and $T_1^k$, $T_2^k$ represent when the vibrations happen. In addition, $n_{abr}(t)$ represents the abrupt noises (e.g., RBM) and $n_{con}(t)$ describes many other constant noises that affect the signal-to-noise ratio (SNR), such as thermal noise~\cite{dong2024robust,lin2024data}, monitoring from random directions~\cite{liu2024diversity} and long-range monitoring~\cite{shen2018respiration}.

\subsubsection{Model of Domain Transformation}
The radar signal modeled in (\ref{equ:vib_long}) shares a strong temporal consistency with the ECG signal as shown in Figure~\ref{fig:scg_ecg}, because the excitation-contraction coupling indicates that the electrical signal (ECG) triggers the heart muscle contraction (SCG)~\cite{swift2021stop}. Therefore, this work proposed to deconstruct the radar-based ECG recovery into three tasks to realize the robust transformation from the measured radar signal $x(t)$ to the ECG signal, and the three tasks can be modeled as:
\begin{itemize}
\item Task $1$: The reconstruction of the morphological features aims to map the single-cycle cardiac activities $x(t)$ to ECG with the deep neural network acting as a mapping function as $x_{ecg}(t) = \mathcal{T}(x(t))$.
\item Task $2$: The detection of R peaks (anchors) is equivalent to finding $\mathbf{R} = \{T^1_1,T^2_1,\cdots,T^K_1\}$ in (\ref{equ:t1}) according to the central frequency $f^k_1$ of $v^k_1$, as shown in Figure~\ref{fig:scg_ecg}.
\item Task $3$: The prediction of the cardiac cycle length is equivalent to finding the peak-to-peak interval (PPI) used for resizing $x_{ecg}$ obtained in Task $1$, as shown in Figure~\ref{fig:scg_ecg}. 
\end{itemize}
Theoretically, PPI can be directly obtained from $\mathbf{R}$ as $T_1^{k+1} - T_1^{k}$, but it is necessary to reckon the PPI estimation to be an individual task in practice, because if one R peak fails to be detected in $\mathbf{R}$, the resultant PPI will be extremely large, destroying the long-term ECG recovery.
\subsubsection{Problem Statement for Domain Transformation}
\begin{figure}[tb]
        \centering
        \subfloat[]{\label{fig:distorted_ecg}\includegraphics[width=0.5\columnwidth]{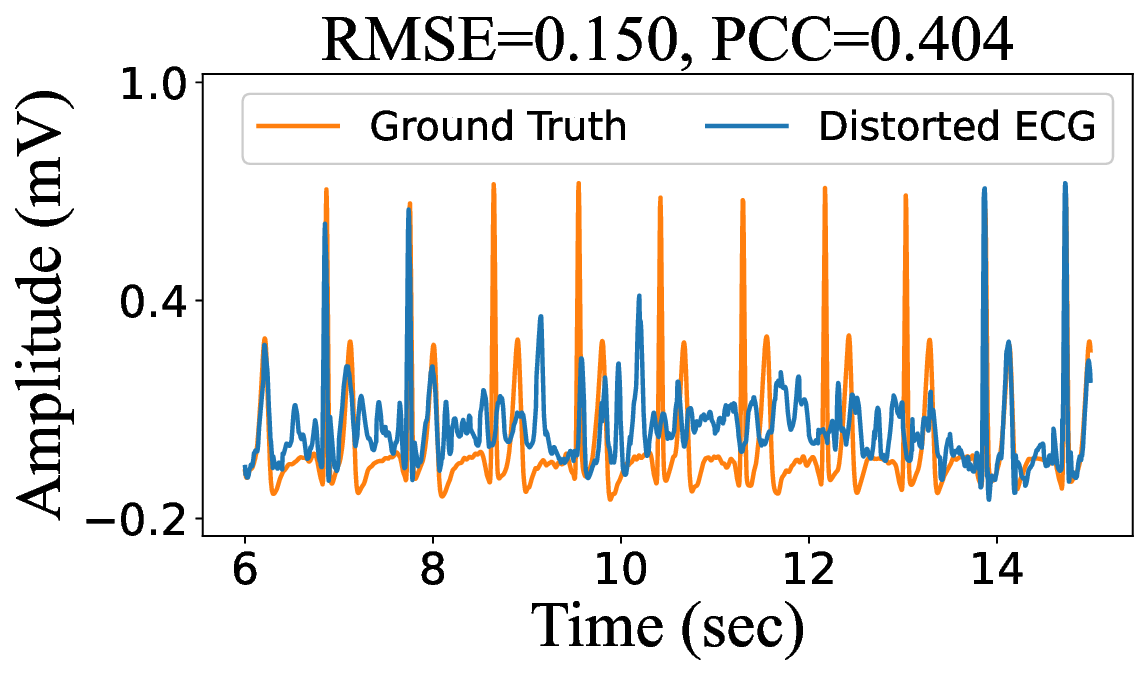}}
        \subfloat[]{\label{fig:mis_ecg}\includegraphics[width=0.5\columnwidth]{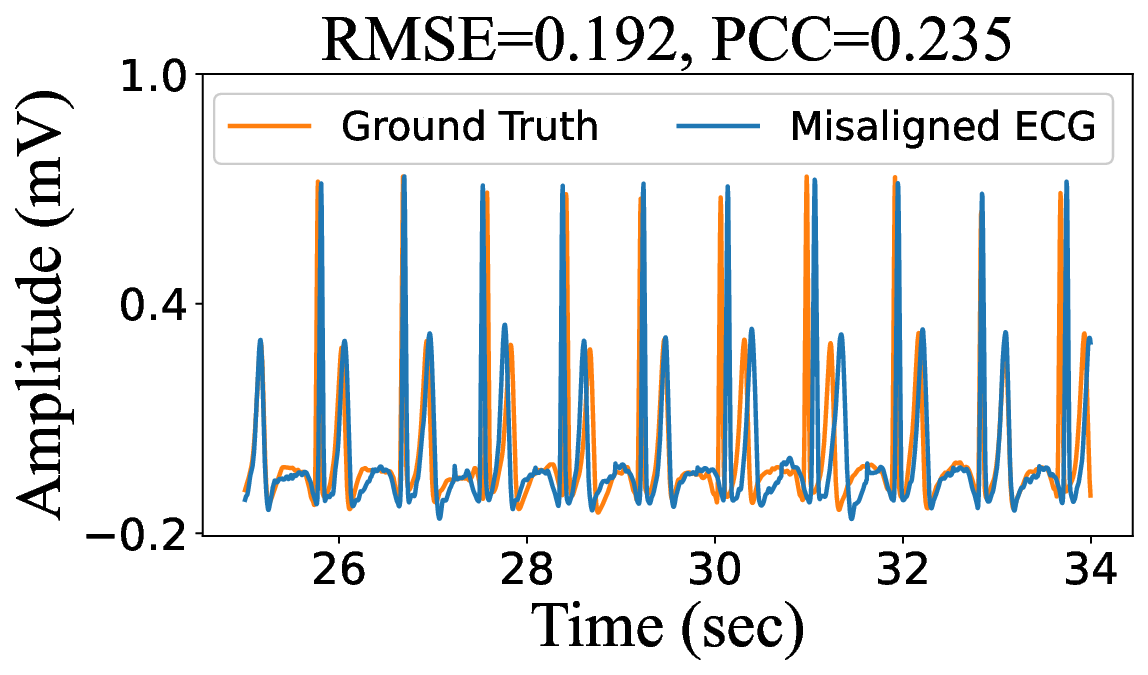}}
        \caption{The impact of strong noise and misalignment: (a) ECG recovery distorted by RBM noise~\cite{chen2022contactless}; (b) Misaligned ECG recovery due to the inaccurate PPI estimation~\cite{zhang2024radarODE}.}
        \label{fig:impact}
\end{figure}

The main problem in the existing domain transformation methods can be summarized as follows:
\begin{itemize}
  \item The transformation between arbitrary radar/ECG pairs is hard to model, and hence the ECG recovery process is vulnerable to the noises with bad root mean square error (RMSE) and Pearson correlation coefficient (PCC) as shown in Figure~\subref*{fig:distorted_ecg}. 
  \item  Although the model for the domain transformation between single-cycle radar/ECG pair has been proposed in~\cite{zhang2024radarODE}, the long-term ECG recovery might be misaligned with ground truth due to inaccurate PPI estimation~\cite{zhang2024radarODE}, deteriorating the RMSE/PCC even if the morphological features are well-recovered as shown in Figure~\subref*{fig:mis_ecg}.
\end{itemize}

In addition, the fine-grained ECG recovery could only realized by deep-learning methods, and the noise robustness of the deep-learning model has never been evaluated in the literature~\cite{chen2022contactless,wang2023ecg,wu2023contactless,zhang2024radarODE}. Therefore, radarODE-MTL dissects the long-term ECG recovery into three tasks, and hence each decoder only focuses on extracting the cardiac feature with different granularity, aiming to improve the accuracy and noise robustness of the radar-based ECG recovery.

\subsection{Optimization Strategies for MTL}\label{sec:rel_mtl}
\subsubsection{Optimization of MTL Network}
A standard definition for an MTL optimization problem with $n$ tasks under hard parameter sharing (HPS~\cite{caruana1993multitask}) architecture is given by:
\begin{equation}\label{equ:mtl_obj}
\boldsymbol{\theta}^*=\underset{\boldsymbol{\theta} \in \mathbb{R}^m}{\arg \min }\left\{\mathcal{F}(\boldsymbol{\theta}) \triangleq \frac{1}{n} \sum_{i=1}^n \mathcal{L}_i(\boldsymbol{\theta})\right\}
\end{equation}
where $\boldsymbol{\theta}\in \mathbb{R}^m$ denotes the shared parameter space, $\mathcal{L}_i(\boldsymbol{\theta})$ is the task-specific non-negative objective function for $\mathbb{R}^m \rightarrow \mathbb{R}_+$, and $\mathcal{F}(\boldsymbol{\theta})$ represents a mapping from the parameter space to the objective space as $\mathbb{R}^m \rightarrow \mathbb{R}^n$. The MTL optimization strategy aims to find the optimal parameter set $\boldsymbol{\theta}^*$ that minimizes the average loss.

The dilemma in the design of MTL optimization strategies is mainly on avoiding negative transfer when the optimization of individual tasks conflicts with each other~\cite{senushkin2023independent,fernando2023mitigating,liu2021towards,liu2021conflict,linsmooth,liu2019end,chennupati2019multinet++,kendall2018multi}, spawning two main categories of methods, loss balancing method and gradient balancing methods, to impartially search for the optimal solution(s) subjecting to Pareto optimality~\cite{liu2021conflict}.

The loss balancing methods add the weight to each task loss $\mathcal{L}_i(\boldsymbol{\theta})$ based on various criteria, such as learning rate~\cite{liu2019end}, inherent task uncertainty~\cite{kendall2018multi} or the loss magnitude~\cite{liu2021towards}. In contrast, gradient balancing methods address the negative transfer by balancing both magnitudes and the directions of the task-specific gradient $\boldsymbol{g}_i = \nabla_{\boldsymbol{\theta}} \mathcal{L}_i(\boldsymbol{\theta})$, according to certain criteria such as the cosine similarity between gradients~\cite{liu2021conflict}, descending rate~\cite{liu2021conflict} or the orthogonality of the gradient system~\cite{senushkin2023independent}.

\subsubsection{Problem Statement for Designing MTL Optimization Strategies}
The existing methods perform not well on the proposed radarODE-MTL framework because most methods aim to treat all the tasks equally and pay too much attention to the easy tasks with the least achievement after convergence (e.g., slow learning rate in GradNorm~\cite{chen2018gradnorm}, small singular value in Aligned-MTL~\cite{senushkin2023independent}), while the hard task tolerates a slow convergence rate due to the limited gradient magnitudes or update frequencies~\cite{guo2018dynamic}. Several studies in the literature proposed to increase the weight for the hard task metered the learning rate~\cite{guo2018dynamic}. However, the forcible change of the weight may aggravate the gradient conflict and hence degrade other tasks, because the loss-balancing method can not alleviate the gradient conflict issue~\cite{liu2021conflict}. 

In addition, the slow learning rate can be interpreted in two ways: (a) The optimization stalls due to the compromise in gradients normalization, and the constraint on the hard task should be released as adopted in GradNorm~\cite{chen2018gradnorm} and DWA~\cite{liu2019end}; (b) The optimization has already achieved convergence and should be terminated as in the early stop technique~\cite{yao2007early}. Unfortunately, it is hardly investigated whether the optimization actually converges or stalls, or say, should more computational resources be skewed towards the task with limited learning progress. Therefore, EGA is proposed in this paper to estimate the intrinsic task difficulty based on the current learning progress and dynamically alter the gradients in orthogonal space to fairly benefit all the tasks without knowing the actual optimization status (i.e., stall or convergence).

\section{Methodology}\label{sec:method}
\subsection{Overview of radarODE-MTL with EGA Strategy}
\begin{figure*}[tb]  
    \centering 
    \includegraphics[width=2\columnwidth]{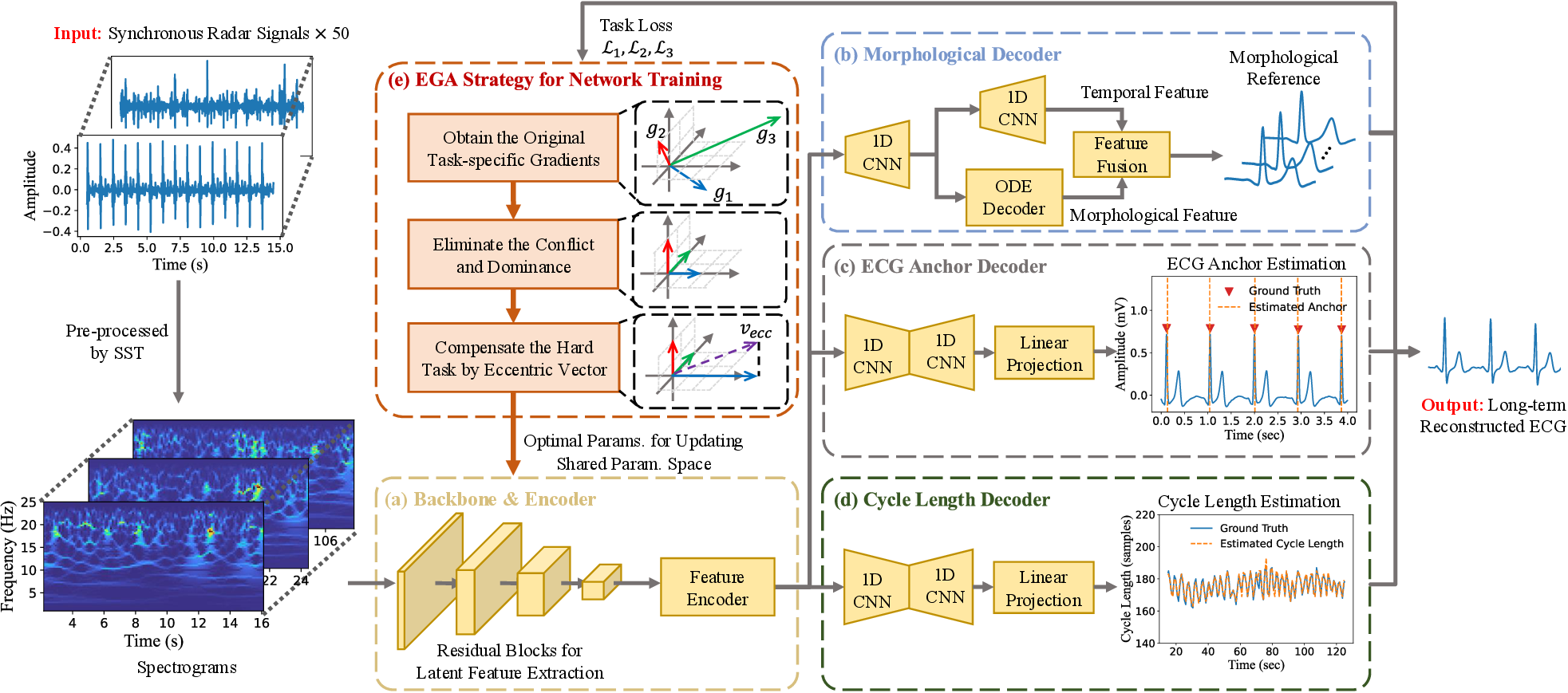}
    \caption{Overview of the radarODE-MTL framework with EGA strategy: (a) Shared backbone extracts time-frequency features from the signal spectrograms with four layers of residual block; (b) Morphological decoder only reconstructs the shape of the current ECG piece; (c) ECG anchor decoder estimates the time-index of anchors (R peaks); (d) Cycle length decoder estimates the length of the current cardiac cycle; (e) The proposed EGA strategy for optimizing shared parameter space.}
    \label{fig:radarODE_mtl} 
\end{figure*}

The aforementioned three deconstructed tasks for radar-based ECG recovery can be realized by the proposed radarODE-MTL framework as shown in Figure~\ref{fig:radarODE_mtl}, and the dataset used for training and validation is provided in~\cite{chen2022contactless}. Firstly, the $50$ synchronous radar signals will be pre-processed into spectrograms by synchrosqueezed transform (SST) to highlight the central frequencies for locating the prominent vibrations $v_1$ and $v_2$. Then, radarODE-MTL is designed to generate the long-term ECG recovery in an end-to-end manner with certain shared layers to capture the common representations for all tasks and three task-specific decoders to recover the ECG morphological features, detect ECG anchors (R peaks) and estimate single-cardiac-cycle length respectively, as shown in Figure~\ref{fig:radarODE_mtl}(a)-(d).

During the training stage, the network optimization of three decoders follows the standard single-task optimization method, and the share parameter space (Backbone$\&$Encoder) is updated using the proposed EGA strategy based on the task-specific loss $\mathcal{L}_1,\mathcal{L}_2,\mathcal{L}_3$, as shown in Figure~\ref{fig:radarODE_mtl}(e). In general, the EGA strategy first tries to eliminate the conflict and dominance among the original task-specific gradients, e.g., $\boldsymbol{g_1}$, $\boldsymbol{g_2}$ have opposite directions and $\boldsymbol{g_3}$ has large magnitude. Secondly, the eccentric vector ($v_{ecc}$) is introduced for balancing the task difficulties to fairly optimize all the tasks.

\begin{remark}
The latent information needed in different tasks can be broadcasted across layers to improve the generalization of the model and the performance of every single task~\cite{lin2023libmtl,senushkin2023independent}. Therefore, in addition to the design of optimization strategies, challenges also arise to designing the efficient MTL structure for knowledge sharing that benefits all the tasks~\cite{liu2019end}.
\end{remark}

\subsection{Backbone and Encoder}
The backbone of radarODE-MTL is used to extract the latent features from the input SST spectrograms as shown in Figure~\ref{fig:radarODE_mtl}(a) and is expected to figure out the remarkable patterns for vibrations $v_1$ and $v_2$ with certain central frequencies and periodicity. Specifically, four residual blocks are adopted in this work as the backbone because the ResNet has been proven to be an efficient structure in computer vision or signal processing~\cite{chen2024tfpred,wang2023fss,chu2024vessel}. Then, the encoder contains only one 2D convolutional layer to further compress the feature in the time-frequency domain into the 1D time domain for later processing. The performance of the backbone and encoder has been verified in our previous work with the detailed structure shown in~\cite{zhang2024radarODE}.

\subsection{Morphological Decoder}
The morphological decoder has been designed in our previous work radarODE~\cite{zhang2024radarODE} as the single cycle ECG generate (SCEG) module to realize the robust domain transformation in a single cardiac cycle with a fast rate of convergence, because an ODE model is introduced in the ODE decoder to provide morphological feature as the prior knowledge to guide/constrain the ECG recovery. Similarly, in radarODE-MTL, a morphological decoder will be used to realize the mapping function $\mathcal{T}(\cdot)$ in Task $1$ and generate morphological reference by fusing both temporal and morphological features, as shown in Figure~\ref{fig:radarODE_mtl}(b).

\subsection{ECG Anchor Decoder and Cycle Length Decoder}
The ECG anchor decoder and cycle length decoder are designed to identify the time-domain anchors $T_1^k$ and single-cardiac-cycle length $PPI^k$ in Task $2$ and $3$ simultaneously for the accurate alignment of ECG pieces as shown in Figure~\ref{fig:radarODE_mtl}(c) and (d), avoiding the impact of error accumulation in long-term ECG recovery~\cite{zhang2024radarODE}. In addition, the prediction of the ECG anchors and cycle lengths can leverage the context information even if the current cardiac cycle is ruined by noises, because the vital signs are nearly unchanged for healthy people in successive cardiac cycles~\cite{xia2021radar}.

The structures of the ECG anchor decoder and cycle length decoder are the same as shown in Figure~\ref{fig:radarODE_mtl}(c) and (d), with several layers of 1D CNN-based encoder/decoder followed by a linear projection block. Specifically, the encoder is assembled by four 1D CNN blocks with each block containing 1D convolution, batch normalization (BN) and rectified linear unit (ReLU) activation function; the decoder is composed of two 1D transposed CNN blocks with each block containing 1D transposed convolution, BN and ReLu; and the linear projection block is assembled by linear layer, BN and ReLU with one linear layer appended at last as the output layer.

\subsection{Input, Output and Loss Function}
The inputs of radarODE-MTL are the $4$-sec segments divided from long-term radar signal with a step length of $1$ sec, and the middle cardiac cycle is selected as the ground truth ECG piece. Then, to calculate the loss value, the ground truth ECG piece should be resampled as a fixed length $200$ to match the output dimension, and the RMSE is used to calculate $\mathcal{L}_1$. The output of the ECG anchor decoder should contain multiple predicted anchors within $4$-sec segment, and the cross-entropy loss is used for $\mathcal{L}_2$ calculation as a multi-class classification problem (i.e., each time index acts as a possible class). Differently, the output of the cycle length decoder only represents the length of the current evaluated cardiac cycle with only one true label (value = $1$), and the cross-entropy loss is used for $\mathcal{L}_3$ calculation as a one-class classification problem.

Eventually, the calculated $\mathcal{L}_1,\mathcal{L}_2,\mathcal{L}_3$ will be used for optimization using the later proposed EGA strategy during training, otherwise the three outputs can directly form the long-term ECG recovery by aligning the recovered ECG pieces (Task $1$) with the predicted anchors (Task $2$) after resampling the ECG pieces as the cycle lengths (Task $3$).
\begin{figure*}[tb]
        \centering
        \subfloat[]{\label{fig:grad_conflict}\includegraphics[width=0.6\columnwidth]{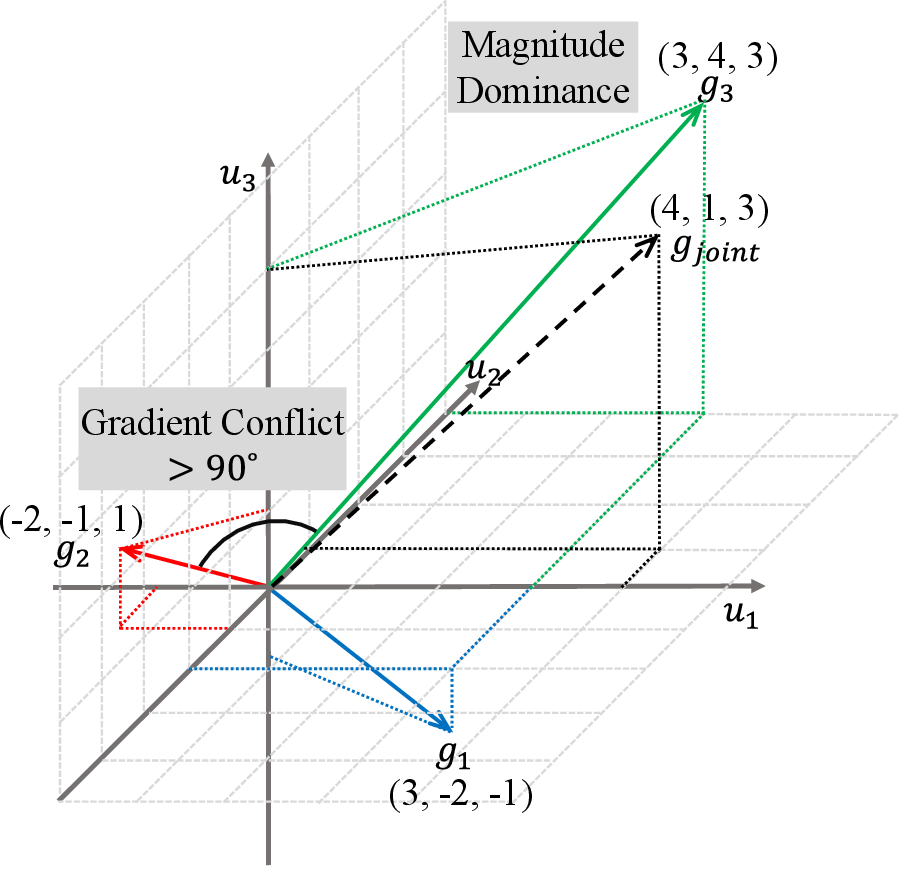}}
        \subfloat[]{\label{fig:grad_orth}\includegraphics[width=0.6\columnwidth]{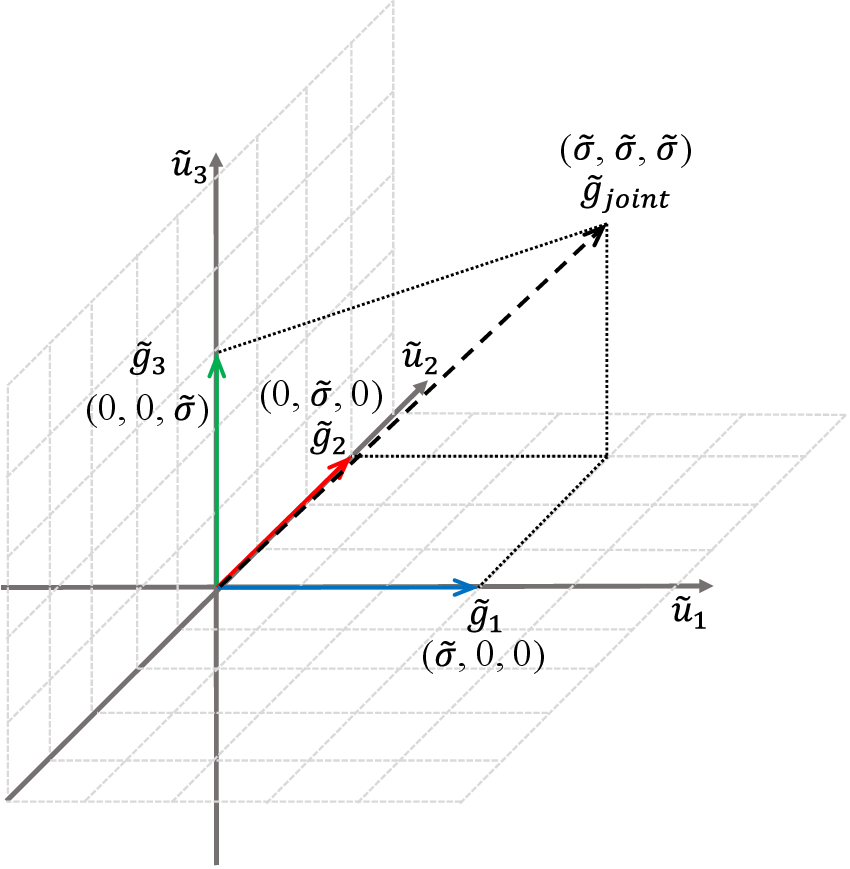}}
        \subfloat[]{\label{fig:grad_ecc}\includegraphics[width=0.6\columnwidth]{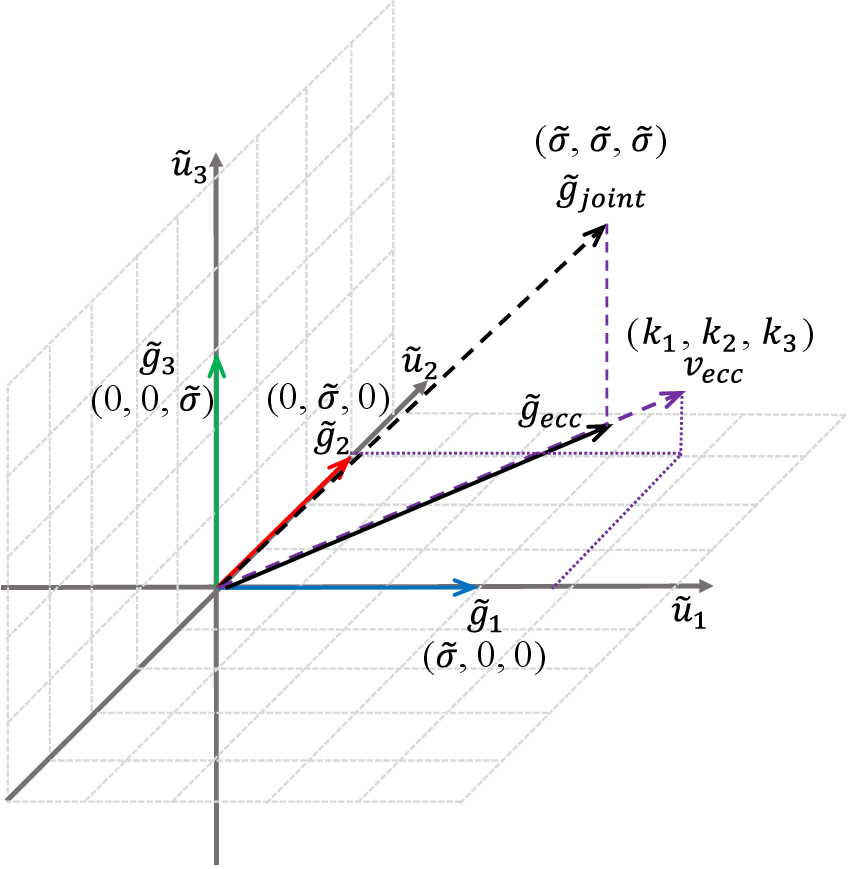}}
        \caption{Illustration of EGA: (a) Original gradient space with gradient conflict and magnitude dominance; (b) The projection of the original gradient space into the orthogonal space with equal “learning rate”; (c) The implementation of eccentric gradient alignment to skew the joint gradient $\tilde{g}_{joint}$ towards the hard task by introducing the eccentric vector $v_{ecc}$.}
        \label{fig:ega}
\end{figure*}

\subsection{Eccentric Gradient Alignment (EGA) Strategy}
According to the discussion in Section~\ref{sec:rel_mtl}, the imbalanced difficulties among three tasks will raise a new challenge to not only simultaneously optimize all the tasks without negative transfer~\cite{linsmooth}, but also keep improving the hard tasks even if the easy tasks have already achieved convergence. 

In this case, EGA first needs to solve the gradient conflict and magnitude dominance within the original task-specific gradients $\boldsymbol{g}_1$, $\boldsymbol{g}_2$, $\boldsymbol{g}_3$ as shown in Figure~\subref*{fig:grad_conflict}, e.g., $\boldsymbol{g}_1$ and $\boldsymbol{g}_2$ may have opposite directions hence canceling with each other, and $\boldsymbol{g}_3$ may have a large magnitude hence dominating the linear combination of all the gradients, with the resultant $\boldsymbol{g}_{joint}$ leaning on $\boldsymbol{g}_3$. A common solution is to project all the gradients into an orthogonal space to eliminate gradient conflict~\cite{senushkin2023independent,dong2022gdod}, and hence the optimization based on $\boldsymbol{g}_{joint}$ will not degrade any of the tasks. Then, the magnitude of the gradients will be unified as the same value (e.g., $\tilde{\sigma}$) to obtain new task-specific gradients $\tilde{\boldsymbol{g}}_{1}$, $\tilde{\boldsymbol{g}}_{2}$, $\tilde{\boldsymbol{g}}_{3}$, as shown in Figure~\subref*{fig:grad_orth}.

Furthermore, instead of categorically selecting the hard task based on the learning rate and only increasing the corresponding weight, EGA creatively provides an adjustable estimation of the intrinsic task difficulty by mapping the learning rate through a softmax with hyperparameter $T$. In other words, suitable intrinsic task difficulty can be obtained by adjusting $T$ without knowing the actual optimization status (i.e., stall or convergence), and the discrepancy among task difficulties can be adjusted to avoid overlooking or overrating any task. In practice, to integrate the estimated intrinsic task difficulty with MTL optimization, EGA proposed to add an eccentric vector $\boldsymbol{v}_{ecc}$ to eccentrically align the joint gradient $\tilde{\boldsymbol{g}}_{joint}$ to the hard task, as shown in Figure~\subref*{fig:grad_ecc}. 

The detailed EGA strategy will be explained in this section in terms of the preparation stage, gradient projection and normalization, and eccentric gradient alignment.

\subsubsection{Preparations for EGA Optimization}
As a gradient-based MTL optimization method with objective function in (\ref{equ:mtl_obj}), EGA requires to access task-specific gradient in terms of the shared parameters $\boldsymbol{\theta}$, and the gradients can be obtained as $\boldsymbol{g}_i = \nabla_{\boldsymbol{\theta}} \mathcal{L}_i(\boldsymbol{\theta}), i\in [n]$, forming the original gradient matrix as $\boldsymbol{G}=\{\boldsymbol{g}_1,\cdots,\boldsymbol{g}_i\}\in \mathbb{R}^{n\times m}$. Then, the joint gradient for optimizing the shared parameter space can be linearly combined as $\boldsymbol{g}_{joint}=\boldsymbol{G^{\intercal}w}$, with $\boldsymbol{w}=[1,\cdots,1]^{\intercal}$ representing the weights for each $\boldsymbol{g}_i$. The original gradient matrix $\boldsymbol{G}$ normally has gradient conflict and magnitude dominance issues, as shown in Figure~\subref*{fig:grad_conflict}.

\subsubsection{Gradients Projection and Normalization} 
In order to solve the conflict inside the gradient matrix $\boldsymbol{G}$, the orthogonal projection problem can be formulated as finding a gradient matrix $\boldsymbol{\tilde{G}}$ with the new joint gradient $\boldsymbol{\tilde{g}}_{joint}=\boldsymbol{\tilde{G}^{\intercal}w}$ close to the original $\boldsymbol{g}_{joint}$:
\begin{equation}
\min\|\boldsymbol{g}_{joint}-\boldsymbol{\tilde{g}}_{joint}\|_2^2 \quad \text { s.t. } \ \tilde{\boldsymbol{G}} \tilde{\boldsymbol{G}}^{\intercal}=\boldsymbol{I}
\end{equation}
Then, according to the derivation based on triangle inequality:
\begin{equation}
\left\|\boldsymbol{g}_{joint}-\tilde{\boldsymbol{g}}_{joint}\right\|_2^2 = \left\|\boldsymbol{G^{\intercal}w}-{\boldsymbol{\tilde{G}^{\intercal}w}}\right\|_2^2 \leq\|\boldsymbol{G^{\intercal}}-{\boldsymbol{\tilde{G}^{\intercal}}}\|_F^2\|\boldsymbol{w}\|_2^2
\end{equation}
At last, the projection problem can be finally formulated as:
\begin{equation}\label{equ:GG}
\min _{\tilde{\boldsymbol{G}}}\|\boldsymbol{G}-\tilde{\boldsymbol{G}}\|_F^2 \quad \text { s.t. } \ \tilde{\boldsymbol{G}} \tilde{\boldsymbol{G}}^{\intercal}=\boldsymbol{I}
\end{equation}

The solution to the problem in (\ref{equ:GG}) has been given in the orthogonal Procrustes problem~\cite{schonemann1966generalized} by simply applying singular value decomposition (SVD) to $\boldsymbol{G}$ as:
\begin{equation}\label{equ:svd}
\boldsymbol{G} = \boldsymbol{U} \boldsymbol{\Sigma} \boldsymbol{V}^{\intercal}
\end{equation}
Then, the orthogonal gradient matrix $\boldsymbol{\tilde{G}}$ with unit singular values can be obtained as:
\begin{equation}\label{equ:sol}
\tilde{\boldsymbol{G}} = \boldsymbol{UV^{\intercal}}
\end{equation}
In addition, the calculation can be simplified by applying the eigenvalue decomposition to the Gram matrices $\boldsymbol{GG^{\intercal}}$ as:
\begin{equation}\label{equ:eign}
\boldsymbol{GG^{\intercal}} = \boldsymbol{U\left(\Sigma \Sigma^{\intercal}\right) U^{\intercal}}
\end{equation}
Then, the final solution in (\ref{equ:sol}) can be rewritten by combining (\ref{equ:svd}) and (\ref{equ:eign}) as:
\begin{equation}\label{equ:orth_unit}
\tilde{\boldsymbol{G}} = \boldsymbol{U\Sigma^{-1} U^{\intercal}G}
\end{equation}

The current $\tilde{\boldsymbol{G}}$ in (\ref{equ:orth_unit}) is orthogonal but with unit singular values, and the next step is to re-scale the task-specific gradients to avoid magnitude dominance. According to the literature~\cite{senushkin2023independent}, the original magnitude of task-specific gradients is proportional to the singular values of $\tilde{\boldsymbol{G}}$. Therefore, to ensure the convergence to the optima of all the tasks, the minimal singular value is selected to calculate the scaling factor instead of using the original singular values, and the re-scaled $\tilde{\boldsymbol{G}}$ can be obtained as:
\begin{equation}\label{equ:G_tilde}
\tilde{\boldsymbol{G}} = \tilde{\sigma}\boldsymbol{U\Sigma^{-1} U^{\intercal}G},\quad \text{with} \ \tilde{\sigma} = \min(\sqrt{\text{eigenvalue}(\boldsymbol{GG^{\intercal}})})
\end{equation}

At last, the orthogonal gradient matrix with equal magnitude is shown in Figure~\subref*{fig:grad_orth}, but all the tasks are currently compromised on the same learning rate, causing the stall of the optimization for certain hard tasks.

\begin{algorithm}[tb]
\caption{EGA Optimization Strategy for MTL}\label{alg:ega}
\begin{algorithmic}[1]
    \State \textbf{Input:} \multiline{0.4}{Loss values for $n$ tasks $[\mathcal{L}_1, \cdots,\mathcal{L}_i], i\in [n]$, \\ Shared parameters $\boldsymbol{\theta}$ and Step length $\eta$, \\ $T$ for softmax and $t_{warm}$ for warmup epoch}
    \State \textbf{Output:} Optimal parameters $\boldsymbol{\theta}^*$ for updating $\boldsymbol{\theta}$
    \Statex \textsc{Objective}:
    \State - Find the optimal parameter set $\boldsymbol{\theta}^*$ such that
    \Statex $\boldsymbol{\theta}^*=\underset{\boldsymbol{\theta} \in \mathbb{R}^m}{\arg \min }\left\{\mathcal{F}(\boldsymbol{\theta}) \triangleq \frac{1}{n} \sum_{i=1}^n \mathcal{L}_i(\boldsymbol{\theta})\right\}$
    % \Statex \textsc{Initialization}:
    \Statex \textsc{For the input batch in certain epoch}:
    \State - Initialize eccentric vector $\boldsymbol{v}_{ecc}=[1,\cdots,1]^{\intercal}\in \mathbb{R}^{n}$
    \State - Get the current epoch as $t$
    \State - Calculate task-specific gradient $\boldsymbol{g}_i = \nabla_{\boldsymbol{\theta}} \mathcal{L}_i(\boldsymbol{\theta}), i\in [n]$
    \State - Form gradient matrix $\boldsymbol{G}=\{\boldsymbol{g}_1,\cdots,\boldsymbol{g}_i\}\in \mathbb{R}^{n\times m}$
    \State - \multiline{0.4}{Calculate eigenvalues/eigenvectors of Gram matrix as in (\ref{equ:eign}):
                  \\$\boldsymbol{GG^{\intercal}} = \boldsymbol{U\left(\Sigma \Sigma^{\intercal}\right) U^{\intercal}}$ with eigenvalues $\boldsymbol{\lambda}$}
    \State - \multiline{0.4}{Get scaling factor: $\tilde{\sigma} = \min(\boldsymbol{\sqrt{\lambda}})$}
    \State - \multiline{0.4}{Calculate the orthogonal and normalized gradient matrix as in (\ref{equ:G_tilde}): $\tilde{\boldsymbol{G}} = \tilde{\sigma}\boldsymbol{U\Sigma^{-1} U^{\intercal}G}$} 
    \If{$t=t_{warm}$}
    \State - Record the loss values for all the tasks $\mathcal{L}_i(t_{warm})$
    % \EndIf
    \ElsIf{$t>t_{warm}$}
    \State - \multiline{0.4}{Calculate the intrinsic task difficulty as in (\ref{equ:diffw}):\\
    $k_i(t) = \text{softmax}(lr_i(t-1))$}
    \State - Form eccentric vector $\boldsymbol{v}_{ecc} = [k_1,\cdots,k_i]^{\intercal}$
    \EndIf
    \State - Calculate final joint gradient $\boldsymbol{\tilde{g}}_{ecc} = \tilde{\boldsymbol{G}}^{\intercal}\boldsymbol{\boldsymbol{v}}_{ecc}$
    \State - Calculate optimal parameters $\boldsymbol{\theta}^*=\boldsymbol{\theta}-\eta \boldsymbol{\tilde{g}}_{ecc}$
\end{algorithmic} 
\end{algorithm} 

\subsubsection{Eccentric Gradient Alignment}
To estimate the intrinsic task difficulty, the first step is to assess the current learning rate $lr_i$ based on the loss value $\mathcal{L}_i$ of each task:
\begin{equation}
lr_i(t-1)=\frac{\mathcal{L}_i(t-1)}{\mathcal{L}_i(t_{warm})}
\end{equation}
with $\mathcal{L}_i(t-1)$ and $\mathcal{L}_i(t_{warm})$ representing the loss value for Task $i$ at the previous epoch and the warmup epoch (e.g., $t_{warm}=4$ in this paper), and the $lr_i$ is inversely proportional to the learning rate (i.e., small $lr_i$ for fast learning rate). Then, a softmax function is applied to mapping the $lr_i$ to the intrinsic task difficulty $k_i$ as:
\begin{equation}\label{equ:diffw}
k_i(t) = \text{softmax}(lr_i(t-1))=\frac{n \exp \left(lr_i(t-1) / T\right)}{\sum_{j=1}^n \exp \left(lr_j(t-1) / T\right)}
\end{equation}
with $T$ controlling the discrepancy of the mapped task difficulties (i.e., small $T$ enlarges the discrepancy between $k_i$), and the summation of the weights should be $\sum_{i=1}^n k_i=n$. In addition, the intrinsic task difficult $k_i$ is positive without the negative transfer issue and can be formed as eccentric vector $\boldsymbol{v}_{ecc} = [k_1,\cdots,k_i]^{\intercal}$ as in Figure~\subref*{fig:grad_ecc} to guide the final joint gradient $\boldsymbol{\tilde{g}}_{ecc}$ for optimization as $\boldsymbol{\tilde{g}}_{ecc} = \tilde{\boldsymbol{G}}^{\intercal}\boldsymbol{\boldsymbol{v}}_{ecc}$. At last, the optimal parameter set $\boldsymbol{\theta}^*$ for updating the shared parameter space can be obtained after providing a step length $\eta$ based on the current parameter set $\boldsymbol{\theta}$ as $\boldsymbol{\theta}^*=\boldsymbol{\theta}-\eta \boldsymbol{\tilde{g}}_{ecc}$.

The entire EGA optimization strategy is summarized in Algorithm~\ref{alg:ega} to repeatedly update the shared parameter space (i.e., Backbone$\&$Encoder in this work) based on all the batches in each epoch, and the optimization will be terminated until achieving a pre-defined epoch number.

\section{Experimental Setting and Result Evaluation}\label{sec:results}
\subsection{Dataset and Implementation}
\subsubsection{Dataset for ECG Recovery}
\begin{figure}[tb] 
    \centering 
    \includegraphics[width=0.9\columnwidth]{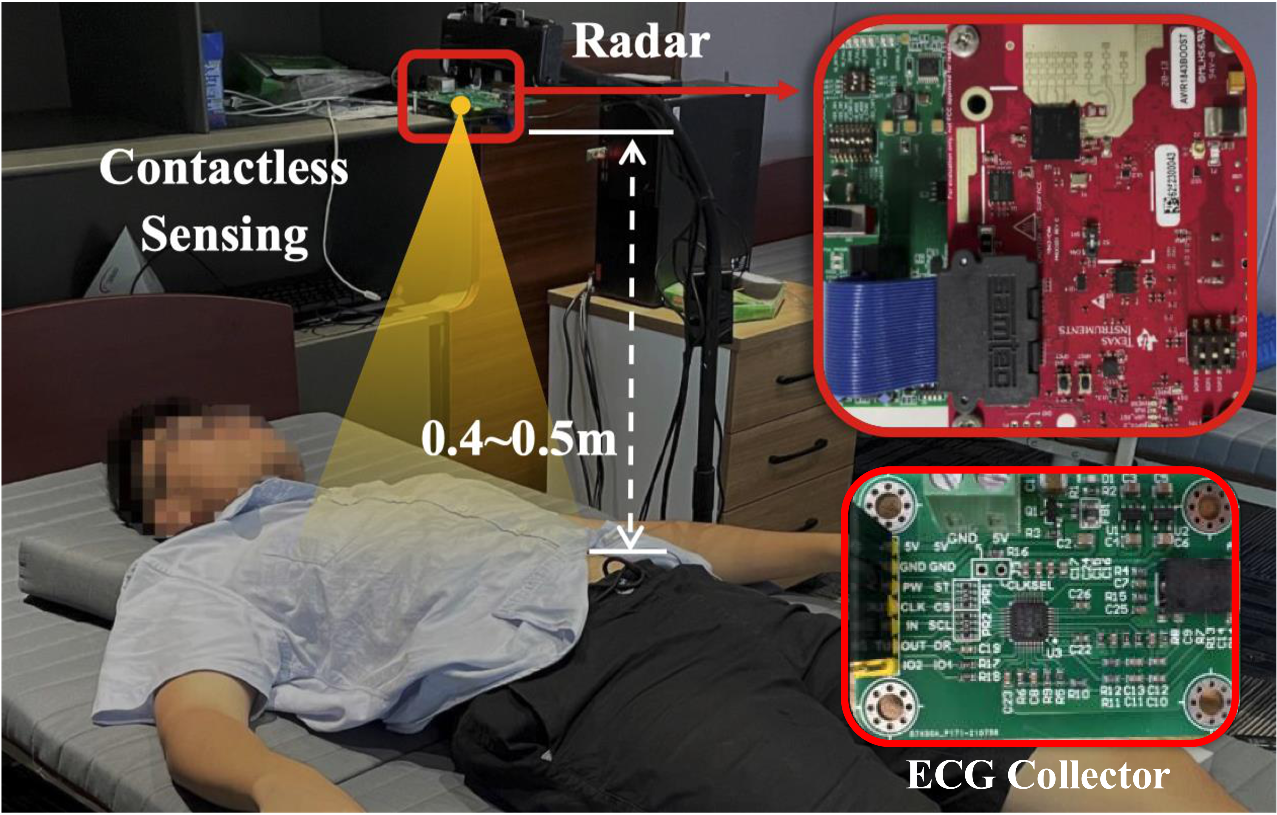}
    \caption{Scenario for data collection from quasi-static subject~\cite{chen2022contactless}.}
    \label{fig:env_set} 
\end{figure}

MMECG~\cite{chen2022contactless} is a dataset used for radar-based ECG recovery and is collected by TI AWR-1843 radar with $77$GHz start frequency and $3.8$GHz bandwidth with the scenario of data collection shown in Figure~\ref{fig:env_set}. A total of $91$ trials for $11$ subjects ($8$ males and $3$ females) are included in the dataset, and each trial lasts for $3$ minutes with synchronous ECG/radar signals sampled at $200$Hz. Following the link budget analysis in~\cite{mercuri2019vital,lin2022broadband}, the collected raw radar signals have a good SNR level of 37dB and are enough for later signal processing or model training.

All the subjects are healthy people without knowing diseases (e.g., premature ventricular contractions~\cite{wang2023ecg}) that may change the common ECG patterns and are asked to keep a quasi-static status to ensure good SNR with the least RBM noise. In addition, the ground ECG signal is collected by TI ADS1292 board with AC coupling and integrated right-leg drive (RLD) amplifier to remove potential baseline drift or power-line noise~\cite{cui2024exploring}.

\subsubsection{Dataset for Evaluating EGA}
NYUv2~\cite{silberman2012indoor} is a dataset for indoor scene understanding recorded using the RGB and Depth cameras and has been widely used as a unified task for validating MTL optimization strategies based on the performance of semantic segmentation, depth estimation, and surface normal prediction~\cite{senushkin2023independent,fernando2023mitigating,liu2021towards,liu2021conflict,linsmooth,liu2019end,chennupati2019multinet++,kendall2018multi}.

\subsubsection{Implementation Details}
The proposed radarODE-MTL along with the radarODE~\cite{zhang2024radarODE} and MMECG~\cite{chen2022contactless} are coded using PyTorch and trained on the NVIDIA RTX A4000 ($16$GB) for $200$ epochs with SGD optimizer~\cite{loshchilov2016sgdr}. The hyperparameters used for training are empirically obtained as batch size $32$,  learning rate $5\times 10^{-3}$, weight decay $5\times 10^{-4}$ and momentum $0.937$. The dataset is split based on different subjects, with the trials from $1$ fixed subject for testing and the other $10$ subjects alternatively selected for training or validation (i.e., $11$-fold cross-validation), ensuring to make use of all the possible trials while not involving the testing data in the training phase. At last, the Python package NeuriKit2~\cite{makowski2021neurokit2} is applied to all the evaluations regarding ECG signals, such as the identification of single cardiac cycles, PQRST peaks detection and heart rate estimation.

The deep learning framework used for scene understanding is implemented in~\cite{lin2023libmtl} with many popular MTL optimization strategies embedded for comparison and optimal hyperparameters provided for training. The training is on the same GPU as before with $200$ epochs, batch size $18$, Adam optimizer~\cite{kingma2014adam}, learning rate $10^{-4}$ and weight decay $10^{-5}$.

\subsection{Performance of EGA}
\subsubsection{Radar-based ECG Recovery}
The performance of EGA is evaluated on three tasks in terms of different metrics: RMSE, PCC and coefficient of determination $R^2$ for the recovered single-cycle ECG pieces; absolute PPI Error for the cycle lengths estimation; and absolute Timing Error and missed detected rate (MDR) for the anchors prediction, with the corresponding comparison across other MTL optimization strategies as shown in Table~\ref{tab:mtl_radar}. In addition, all the experiments are repeated five times, and the last column $\Delta m\%$ in Table~\ref{tab:mtl_radar} shows a comprehensive assessment across $n$ tasks with $95\%$ confidence interval (CI) and is calculated as:
\begin{equation}\label{equ:mdr}
\Delta m\%=\frac{1}{n} \sum_{i=1}^{n} \frac{1}{n_i} \sum_{j=1}^{n_i} S_{i, j} \frac{M_{m,i,j}-M_{b, i, j}}{M_{b, i, j}} \times 100\%
\end{equation}
where $n_{i}$ is the number of metrics for task $i$, $M_{m,i,j}$ means the performance of a method $m$ on the task $i$ measured with the metric $j$, $M_{b,i,j}$ represents the performance for the single-task baseline, and $S_{i, j}=1/0$ if lower/higher values are better for the current metric (indicated by $\downarrow / \uparrow$). Lastly, the T-test is adopted with the $P$-value calculated for all the experiments as shown in Table~\ref{tab:mtl_radar}, and the statistical analysis will be given at the end of each subsection.

\begin{table*}[tb]
  \caption{Comparison of different optimization strategies on radar-based ECG recovery}
  \centering
  \begin{tabular}{lcccccc|rc}
  \toprule
  \diagbox{\textbf{Methods}}{\textbf{Tasks}} & \multicolumn{3}{c}{\textbf{\makecell[c]{ECG Shape Recovery}}} & \multicolumn{1}{c}{\textbf{\makecell[c]{Cycle Length \\ Estimation}}} & \multicolumn{2}{c|}{\textbf{\makecell[c]{ECG Anchor Estimation}}} & \multirow{3}*{$\Delta m\% \uparrow$} & \multirow{3}*{\makecell[c]{$P$ Value \\ ($\times 10^{-2}$)}} \\
   & RMSE (mV) $\downarrow$ & PCC $\uparrow$ & $R^2$ $\uparrow$ & \makecell[c]{PPI Error \\ (ms)} $\downarrow$  & \makecell[c]{Timing Error\\ (ms)} $\downarrow$ & MDR $\downarrow$ &  \\
  \midrule
  \textbf{Single-task baseline} & 0.106 & 86.6\% & 0.81 & 9.6 & 7.5 & 6.67\% & 0.00$\pm$1.43 & -\\
  \midrule
  \multicolumn{8}{c}{Loss Balancing Methods} \\
  Equal Weight & 0.125 & 79.7\% & 0.63 & \textbf{8.0} & 9.7 & 5.51\% & -1.78$\pm$2.16 & 9.26 \\
  UW~\cite{kendall2018multi} & \textbf{0.066} & 88.5\% & 0.85 & 11.2 & \textbf{5.5} & 6.44\% & 4.04$\pm$3.79 & 2.43 \\
  GLS~\cite{chennupati2019multinet++} & 0.087 & 87.3\% & 0.81 & 14.1 & 6.7 & 4.32\% & -5.89$\pm$2.02 & 0.02 \\
  DWA~\cite{liu2019end} & 0.133 & 80.7\% & 0.79 & {8.3} & {6.4} & 5.33\% & 6.45$\pm$3.71 & 0.20 \\
  STCH~\cite{linsmooth} & 0.070 & 88.0\% & \underline{0.86} & 13.9 & \textbf{5.5} & \textbf{3.28}\% & 2.90$\pm$3.21 & 5.12 \\
  \midrule
  \multicolumn{8}{c}{Gradient Balancing Methods} \\
  CAGrad~\cite{liu2021conflict} & 0.107 & 84.2\% & 0.79 &10.2 & 6.2 & 3.98\% & 6.84$\pm$2.12 & 0.01 \\
  IMTL~\cite{liu2021towards} & 0.088 & \underline{89.4\%} & \underline{0.86} & 9.3 & \underline{6.0} & 6.22\% & 8.43$\pm$1.39 & 0.00 \\
  MoCo~\cite{fernando2023mitigating} & 0.179 & 61.0\% & 0.66 & 8.7 & 6.8 & 4.27\% & -2.32$\pm$1.37 & 1.16 \\
  Aligned-MTL~\cite{senushkin2023independent} & 0.092 & 87.9\% & 0.84 & 10.0 & 6.9 & {3.52\%} & 10.14$\pm$1.11 & 0.00 \\
  \midrule
  \textbf{EGA ($T=0.1$)} & 0.119 & 79.0\% & 0.72& 10.6 & 6.8 & \underline{3.34\%} & 2.83$\pm$0.98 & 0.19 \\
  \textbf{EGA ($T=0.5$)} & \underline{0.082} & \textbf{89.6\%} & \textbf{0.87}& 9.9 & 6.3 & {4.19\%} & \underline{11.55$\pm$1.44} & 0.00 \\
  \textbf{EGA ($T=1.0$)} & 0.085 & 87.4\% & 0.85 & 8.5 & 7.2 & 4.31\% & \textbf{13.37$\pm$1.36} & 0.00 \\
  \textbf{EGA ($T=1.5$)} & 0.105 & 82.9\% & 0.78 & \underline{8.1} & 6.3 & 5.13\% & 10.94$\pm$1.30 & 0.00 \\
  \textbf{EGA ($T=2.0$)} & 0.091 & 86.3\% & 0.84 & 9.2 & 7.3 & 4.01\% & 10.43$\pm$0.95 & 0.00 \\
  \bottomrule
  \multicolumn{9}{r}{\textbf{Bold} and \underline{underline} represent the best and the second best results, respectively.} \\
  \end{tabular}
  \label{tab:mtl_radar}
\end{table*}

In general, the effect of unbalanced task-specific gradients is revealed by using equal weight as shown in Table~\ref{tab:mtl_radar}, and the performance of the hard task (ECG shape recovery) is much worse than baseline while the PPI error even achieves the best accuracy. After balancing the magnitudes and directions of task-specific gradients, the proposed EGA strategy meets the expectation by adjusting the value of $T$ with the following evaluations:
\begin{itemize}
  \item EGA with $T=1.0$ achieves the largest improvement with $\Delta m\% = 13.37$ but none of the individual metrics gets the best or second-best result, and $T=1.0$ can be viewed as a suitable estimation of intrinsic task difficulty to achieve unbiased improvements across all tasks.
  \item EGA with $T=0.5$ obtains the second-best overall performance with $\Delta m\% = 11.55$ and becomes the best in learning ECG morphological features according to RMSE/PCC/$R^2$, indicating $T=0.5$ slightly overrates the difficulty of Task $1$.
  \item EGA with $T=0.1$ cannot balance the task difficulties, hence getting a low score.
  \item EGA with large $T$ values ($1.5$ and $2.0$) tend to evenly distribute the task difficulty weights, and the performance should be similar to other orthogonality-based methods (e.g., Aligned-MTL).
\end{itemize}

In addition, it is also worth noticing that some methods achieve a significant improvement on a particular task, e.g., UW obtains $\text{RMSE}=0.066$mV and $\text{PPI Error}=5.5$ms, implying a potential improvement probably by enlarging the parameter space (scaling the model size) or designing a more efficient MTL architecture instead of using simple HPS~\cite{jeong2024quantifying}. However, the method with remarkable performance on the single task cannot achieve unified improvement on other tasks, e.g., UW and STCH both get good results in ECG anchor estimation ($\text{Timing Error}=5.5$ms), but a huge degradation happens on the cycle length estimation ($\text{PPI Error}>10$ms), revealing the effectiveness of EGA to avoid overvaluing one certain task.

Lastly, when comparing EGA to the method also based on orthogonality, Aligned-MTL stalls after Task $3$ achieves convergence (low $MDR=3.52\%$), while EGA ($T=1.0$) keeps improving Task $1$ and $2$ and gets a better result on $\text{RMSE}=0.085$mV and $\text{PPI Error}=8.5$ms with only a slight degradation on Task $3$ ($\text{Timing Error}=7.2$ms, $\text{MDR}=4.31\%$), showing the ability of EGA to focus on the hard task without distracted by the well-trained easy tasks.

\begin{table*}[t]
  \centering
  \caption{Comparison of different optimization strategies on indoor scene understanding}
  \label{tab:results}
  \begin{tabular}{lccccccccc|rc}
  \toprule
  \multirow{3}*{\textbf{Method}} & \multicolumn{2}{c}{\multirow{2}*{\textbf{Segmentation $\uparrow$}}} & \multicolumn{2}{c}{\multirow{2}*{\textbf{Depth Estimation $\downarrow$}}} & \multicolumn{5}{c|}{\textbf{Surface Normal Prediction}} & \multirow{3}*{$\Delta m\% \uparrow$} & \multirow{3}*{\makecell[c]{$P$ Value \\ ($\times 10^{-2}$)}} \\
  \multicolumn{5}{c}{} & \multicolumn{2}{c}{Angle Distance $\downarrow$} & \multicolumn{3}{c|}{Within t$^\circ \uparrow$}  \\
   & mIoU & Pixel Acc. & Abs. Err. & Rel. Err. & Mean & Median & 11.25 & 22.5 & 30 \\
  \midrule
  \textbf{Single-task baseline} & 52.08 & 74.11 & 0.4147 & 0.1751 & 23.83 & 17.36 & 34.34 & 60.22 & 71.47 & 0.00$\pm$0.19 & - \\
  \midrule
  \multicolumn{11}{c}{Loss Balancing Methods} \\
  Equal Weight & \textbf{53.36} & 74.94 & 0.3953 & 0.1672 & 24.35 & 17.55 & 34.22 & 59.64 & 70.71 & 1.75$\pm$1.60 & 1.67 \\
  UW~\cite{kendall2018multi} & \underline{53.33} & \textbf{75.43} & \underline{0.3878} & 0.1639 & 24.03 & 17.24 & 34.80 & 60.33 & 71.31 & 2.92$\pm$2.05 & 0.44 \\
  GLS~\cite{chennupati2019multinet++} & 53.04 & 74.68 & 0.3951 & 0.1600 & 24.03 & 17.30 & 34.78 & 60.17 & 71.28 & 2.69$\pm$1.52 & 0.12 \\
  DWA~\cite{liu2019end} & 53.12 & \underline{75.23} & {0.3883} & 0.1615 & 24.26 & 17.60 & 34.25 & 59.51 & 70.62 & 2.55$\pm$1.91 & 0.62 \\
  STCH~\cite{linsmooth} & 52.87 & 74.78 & 0.3915 & 0.1615 & 23.27 & \underline{16.34} & \textbf{36.61} & \underline{62.33} & 72.98 & \textbf{3.99$\pm$0.61} & 0.00 \\
  \midrule
  \multicolumn{11}{c}{Gradient Balancing Methods} \\
  CAGrad~\cite{liu2021conflict} & 52.19 & 74.07 & 0.3976 & 0.1634 & 23.83 & 17.16 & 34.89 & 60.65 & 71.77 & 2.09$\pm$1.11 & 0.09 \\
  IMTL~\cite{liu2021towards} & 52.34 & 74.35 & 0.3897 & \underline{0.1579} & 23.76 & 17.00 & 35.28 & 60.92 & 71.89 & 3.24$\pm$0.78 & 0.00 \\
  MoCo~\cite{fernando2023mitigating} & 52.78 & 74.59 & \textbf{0.3858} & 0.1612 & 23.34 & 16.51 & 36.21 & 61.90 & 72.65 & 3.94$\pm$0.72 & 0.00 \\
  Aligned-MTL~\cite{senushkin2023independent} & 52.19 & 74.17 & 0.3911 & 0.1605 & 23.44 & 16.73 & 35.45 & 61.74 & 72.70 & 3.24$\pm$1.08 & 0.00 \\
  \midrule
  % EGA 
  \textbf{EGA} ($T=0.1$) & 52.16 & 74.23 & 0.3944 & 0.1651 & 23.32 & 16.62 & 35.87 & 61.81 & 72.72 & 2.84$\pm$0.88 & 0.00 \\
  \textbf{EGA} ($T=0.5$) & 51.82 & 73.98 & 0.3904 & 0.1614 & 23.41 & 16.66 & 35.87 & 61.65 & 72.51 & 3.11$\pm$0.51 & 0.00 \\
  \textbf{EGA} ($T=1.0$) & 51.75 & 74.38 & 0.3913 & 0.1609 & \textbf{23.09} & \textbf{16.29} & \underline{36.54}& \textbf{62.51} & \textbf{73.22} & 3.71$\pm$1.08 & 0.00 \\
  \textbf{EGA} ($T=1.5$) & 52.37 & 74.65 & 0.3950 & \textbf{0.1571} & \underline{23.15} & 16.46 & 36.07 & 62.22 & \underline{73.07} & \underline{3.96$\pm$0.98} & 0.00 \\
  \textbf{EGA} ($T=2.0$) & 52.18 & 74.23 & 0.3922 & 0.1605 & 23.28 & 16.61 & 35.77 & 61.95 & 72.86 & 3.39$\pm$1.18 & 0.00 \\
  \bottomrule
  \multicolumn{12}{r}{\textbf{Bold} and \underline{underline} represent the best and the second best results, respectively.} \\
  \end{tabular}
  \label{tab:mtl_nyu}
\end{table*}

\begin{figure*}[tbp]
        \centering
    \begin{minipage}[t]{0.6\columnwidth}
    \subfloat[]{\label{fig:radar_good}\includegraphics[width=1\columnwidth,valign=t]{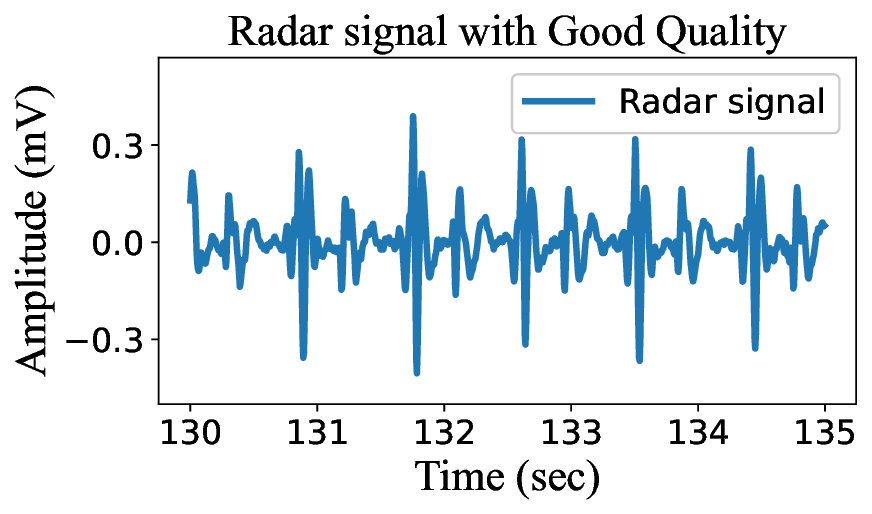}} \\  
    \subfloat[]{\label{fig:mmecg_good}\includegraphics[width=1\columnwidth]{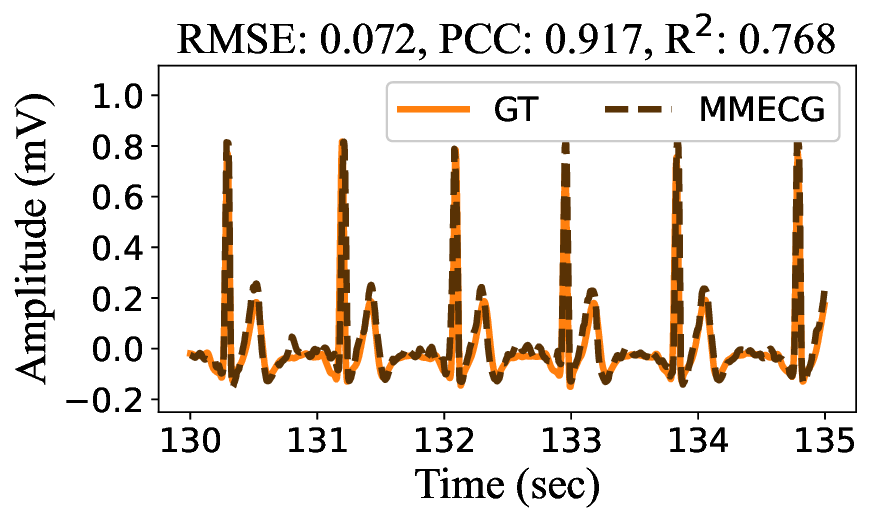}}\\
    \subfloat[]{\label{fig:ode_good}\includegraphics[width=1\columnwidth]{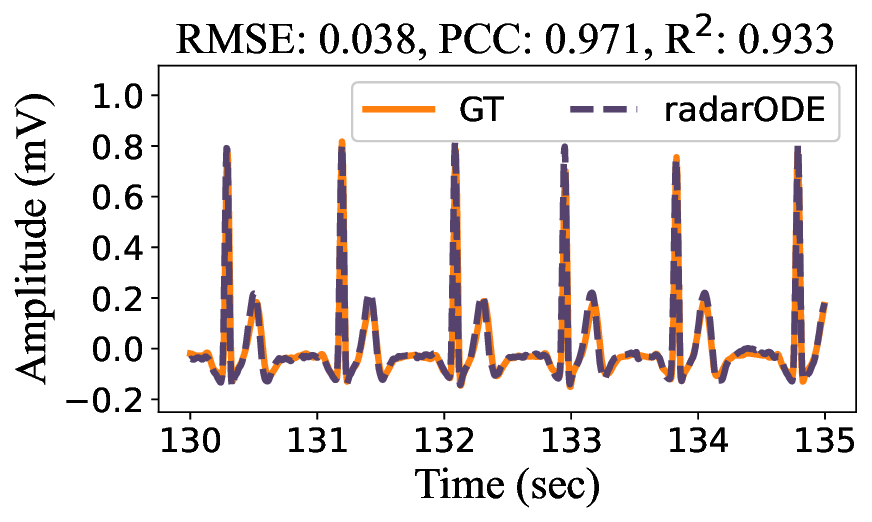}}\\
    \subfloat[]{\label{fig:mtl_good}\includegraphics[width=1\columnwidth]{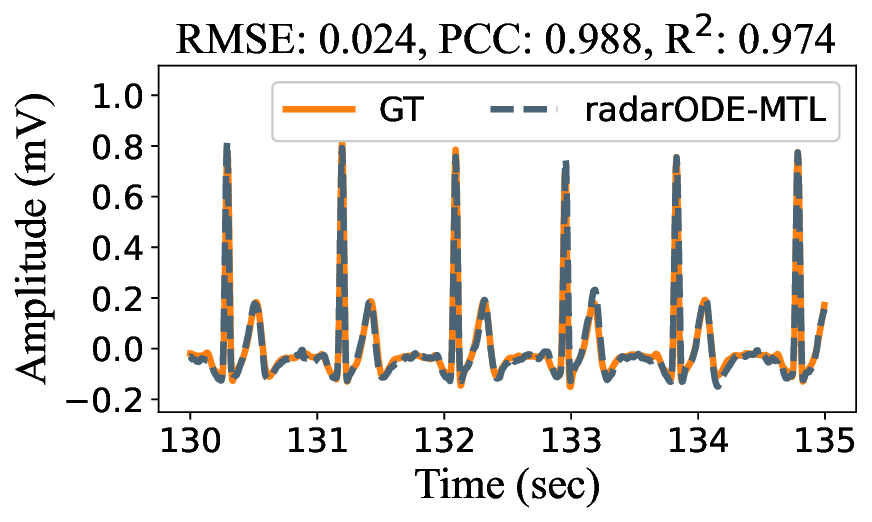}}\\
    \end{minipage}%
    \begin{minipage}[t]{0.6\columnwidth}
    \subfloat[]{\label{fig:radar_moderate}\includegraphics[width=1\columnwidth,valign=t]{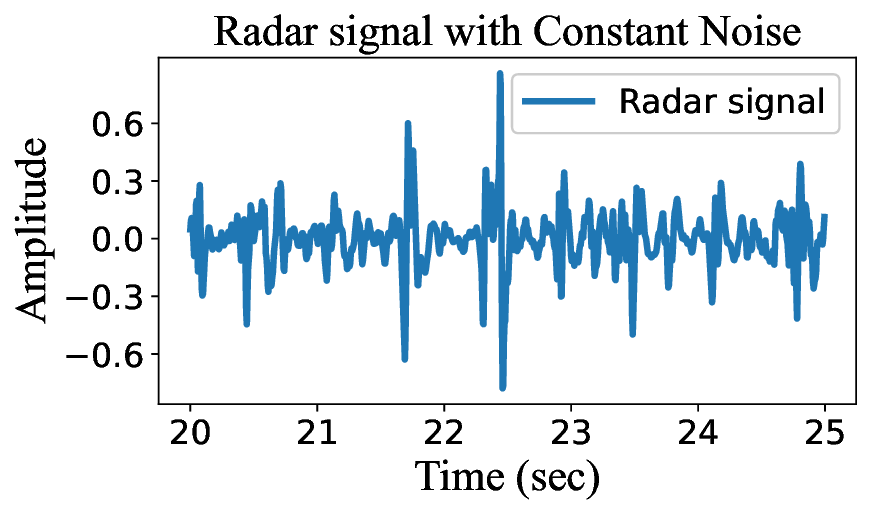}} \\  
    \subfloat[]{\label{fig:mmecg_moderate}\includegraphics[width=1\columnwidth]{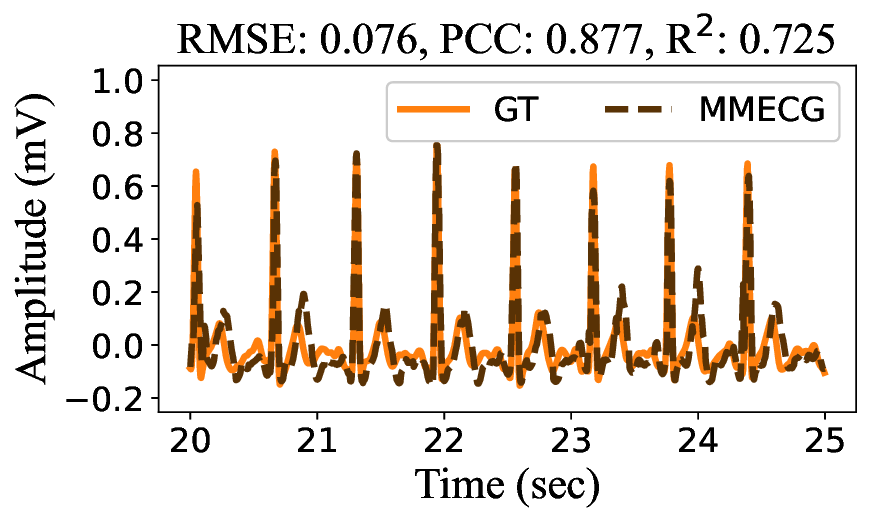}}\\
    \subfloat[]{\label{fig:ode_moderate}\includegraphics[width=1\columnwidth]{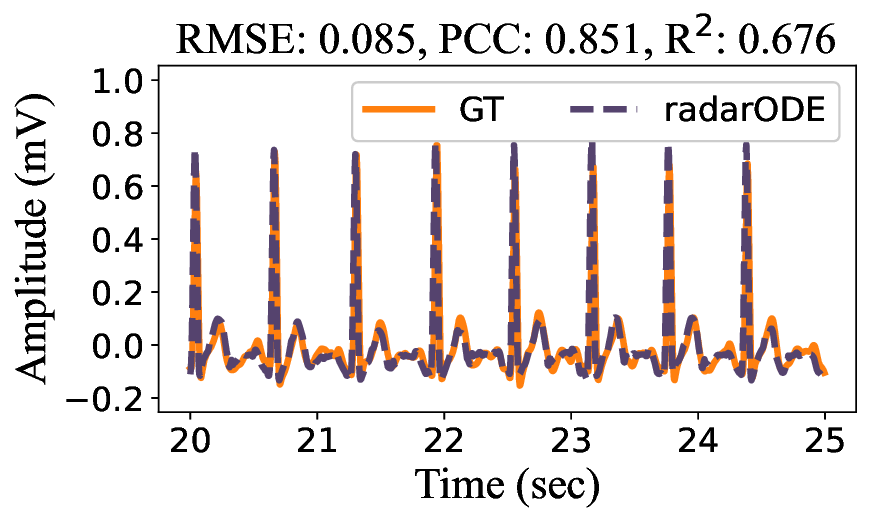}}\\
    \subfloat[]{\label{fig:mtl_moderate}\includegraphics[width=1\columnwidth]{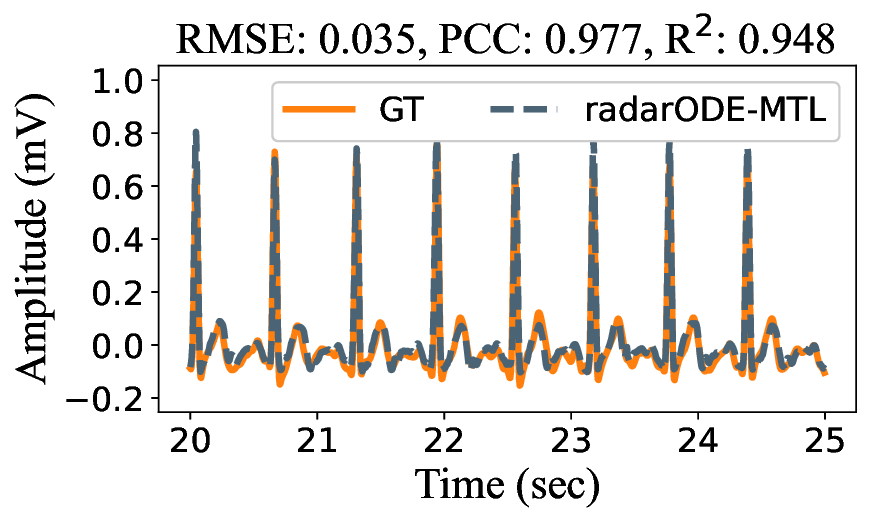}}\\
    \end{minipage}%    
    \begin{minipage}[t]{0.6\columnwidth}
    \subfloat[]{\label{fig:radar_abrupt}\includegraphics[width=1\columnwidth,valign=t]{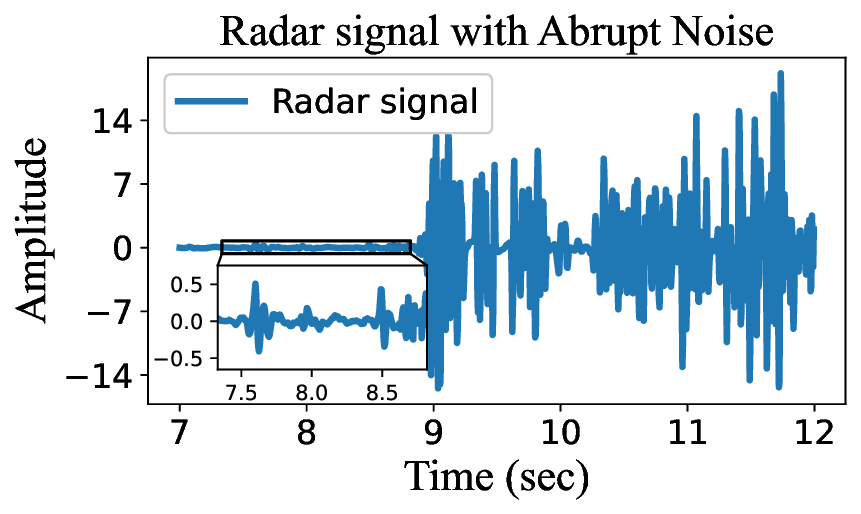}} \\  
    \subfloat[]{\label{fig:mmecg_abrupt}\includegraphics[width=1\columnwidth]{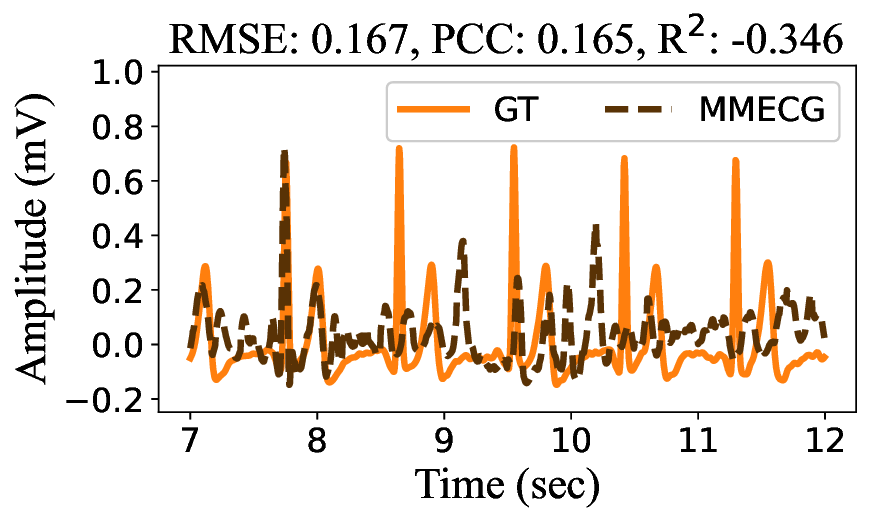}}\\
    \subfloat[]{\label{fig:ode_abrupt}\includegraphics[width=1\columnwidth]{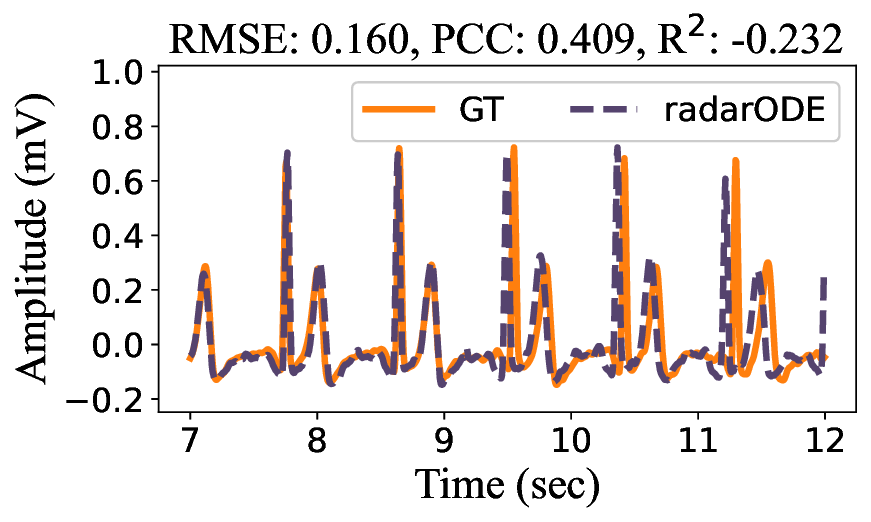}}\\
    \subfloat[]{\label{fig:mtl_abrupt}\includegraphics[width=1\columnwidth]{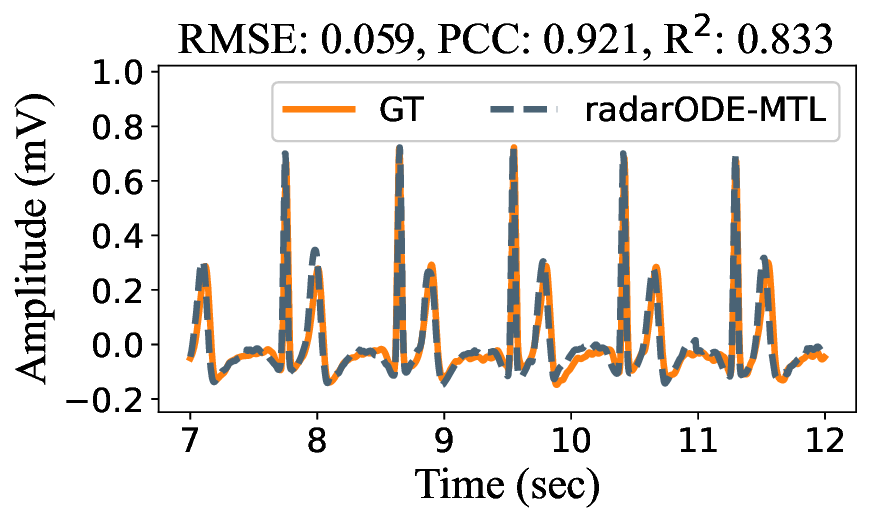}}\\
    \end{minipage}%
    \caption{Illustration of the recovered ECG with ground truth (GT) under different frameworks and noises: (a) - (d) Radar signal with good quality and the recovered ECG signals; (e) - (h) Radar signal with constant noise and the recovered ECG signals; (i) - (l) Radar signal with abrupt noise and the recovered ECG signals. }
    \label{fig:compare_noise}
\end{figure*}
\subsubsection{Indoor Scene Understanding}
The indoor scene understanding based on NYUv2 is a commonly adopted task by all the studies about MTL optimization strategies~\cite{silberman2012indoor}. The metrics for each task are: mean intersection over union (mIoU) and pixel accuracy (Pixel Acc.) for segmentation, absolute/related error (Abs./Rel. Err.) for depth estimation and mean/median angle distance, and the percentage of surface normal within $t^{\circ}$ for surface normal prediction, as shown in the heads of Table~\ref{tab:mtl_nyu}.

According to the improvements $\Delta m\%$ in Table~\ref{tab:mtl_nyu}, EGA ($T=1.5$ and $1.0$) achieves a competitive result compared with other powerful methods, indicating that EGA can be applied to other MTL tasks with an appropriate selection of $T$. An interesting observation is that some methods with average or even poor performance in Table~\ref{tab:mtl_radar} (i.e., STCH and MoCo) achieve remarkable results in scene understanding. A possible explanation is that the indoor scene understanding task may have a small discrepancy in task difficulties and fewer conflicts in gradient directions. This guess can also be verified by the fact that loss balancing methods achieve competitive performance compared with gradient balancing methods, and different $T$ values have limited impacts on the final performance of EGA.

\textbf{Statistical Analysis for EGA: }To verify the significance of the EGA performance, the well-known T-test is performed with the null hypothesis as the performances of the compared methods are identical. The yielded $P$ values for all the methods are shown in Table~\ref{tab:mtl_radar} and~\ref{tab:mtl_nyu}, and all the methods achieve $P<0.05$ except for using the equal weights, indicating that the obtained mean values are reliable with significance. In addition, it is worth noticing that the CIs of the loss balancing methods are larger than that of the gradient balancing methods for both tasks as shown in Table~\ref{tab:mtl_radar} and~\ref{tab:mtl_nyu}, coinciding with the previous conclusion that the access of the gradients is beneficial to the stability of model training. Lastly, the proposed EGA not only achieves the best performance in ECG recovery, but the obtained results are also stable with small CIs, owing to the orthogonal projection that decreases the condition number of the gradient system~\cite{senushkin2023independent}.

To conclude the above evaluations in terms of different tasks, the proposed EGA could successfully alleviate the gradient conflicts and magnitude dominance in MTL optimization, while the intrinsic task difficulty can be successfully estimated to guide the optimization direction by introducing eccentric vector $\boldsymbol{v}_{ecc}$. Compared with other methods, EGA achieves an outstanding result for the tasks with disparate difficulties and is also competitive in the common tasks, but the hyperparameter $T$ should be carefully selected. During practice, the $T$ value can be adjusted until achieving optimum based on the fact that large $T$ evenly treats all the tasks and small $T$ enhances the hard task.

\subsection{Evaluations on the Long-term Recovered ECG}
\subsubsection{General Visualization for ECG Reconstruction}
The outputs from Task $1-3$ can form the long-term ECG signal as depicted in Figure~\ref{fig:compare_noise}. All three frameworks successfully reconstruct the ECG signals from  high-quality radar signals as shown in Figure~\subref*{fig:mmecg_good},~\subref*{fig:ode_good} and~\subref*{fig:mtl_good}, only with certain fluctuations in MMECG result and causing low RMSE/PCC/$R^2$. 

In the presence of constant or abrupt noise, the signal SNR will decrease with the subtle features (e.g., $v_2$) ruined as shown in Figure~\subref*{fig:radar_moderate} and~\subref*{fig:radar_abrupt}. MMECG shows the least noise robustness and cannot resist abrupt noise as also reported in the benchmark paper~\cite{chen2022contactless} as shown in Figure~\subref*{fig:mmecg_moderate} and~\subref*{fig:mmecg_abrupt}, while radarODE achieves robust ECG recovery within each single cardiac cycle but shows obvious misalignment due to the PPI estimation error as shown in Figure~\subref*{fig:ode_moderate} and~\subref*{fig:ode_abrupt}. Lastly, the proposed radarODE-MTL realizes the ECG reconstruction in an end-to-end manner without reintroducing the noises, and the recovered ECG is less corrupted by the noises as shown in Figure~\subref*{fig:mtl_moderate} and~\subref*{fig:mtl_abrupt}.

\begin{figure*}[tb]
        \centering
        \subfloat[]{\label{fig:mdr_cdf}\includegraphics[width=0.41\columnwidth]{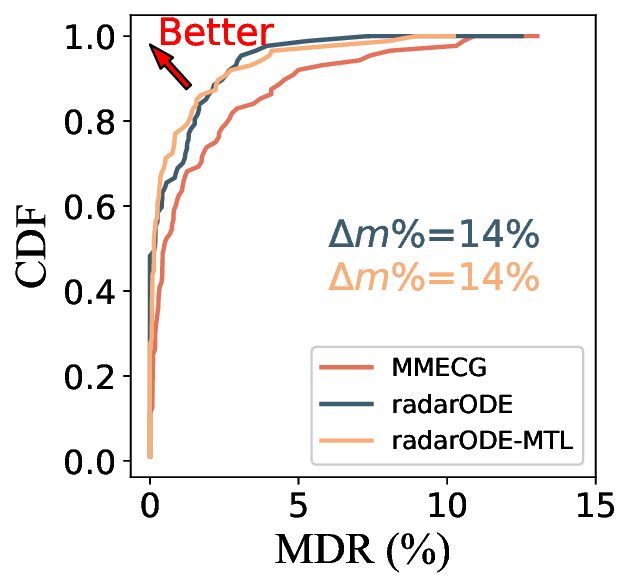}}
        \subfloat[]{\label{fig:hr_cdf}\includegraphics[width=0.39\columnwidth]{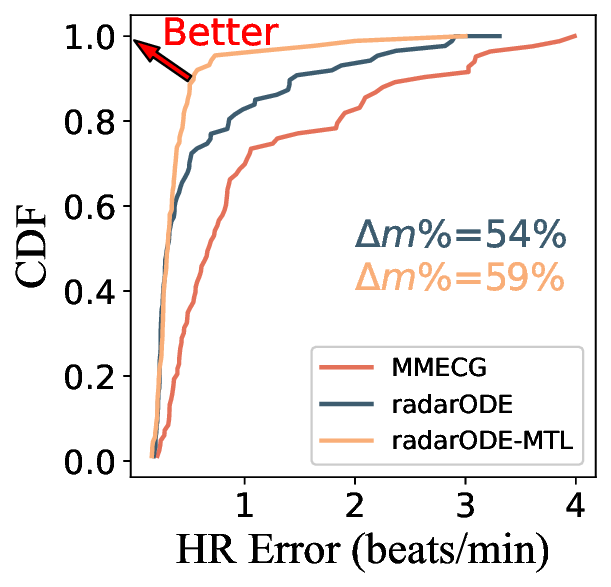}}
        \subfloat[]{\label{fig:rmse_cdf}\includegraphics[width=0.41\columnwidth]{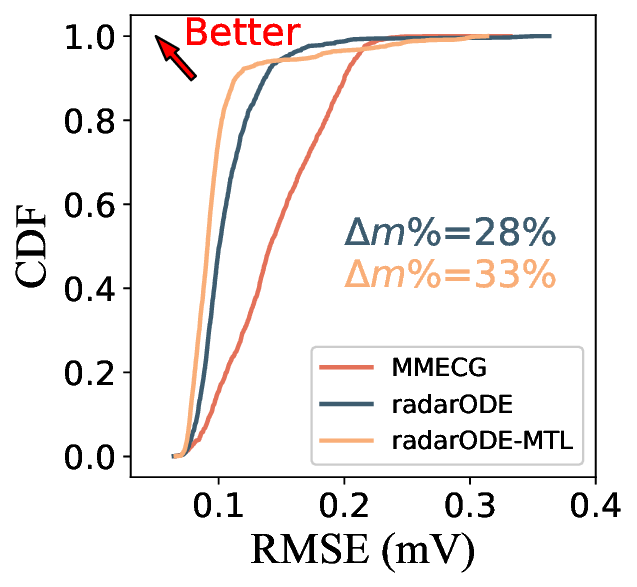}}
        \subfloat[]{\label{fig:ocor_cdf}\includegraphics[width=0.41\columnwidth]{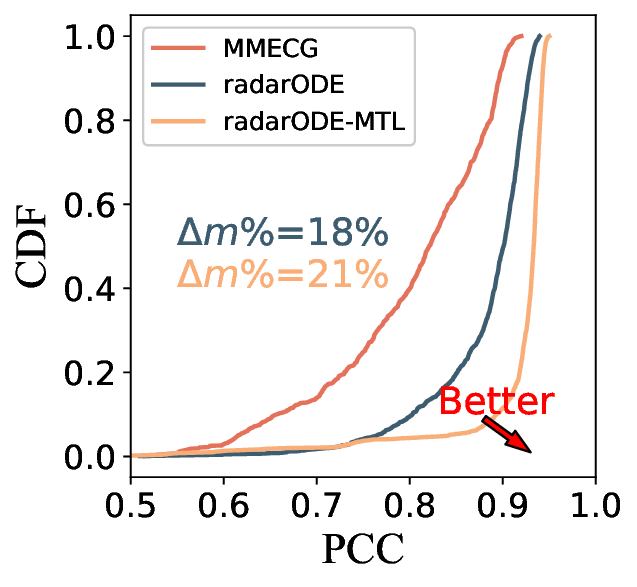}}
        \subfloat[]{\label{fig:r2_cdf}\includegraphics[width=0.41\columnwidth]{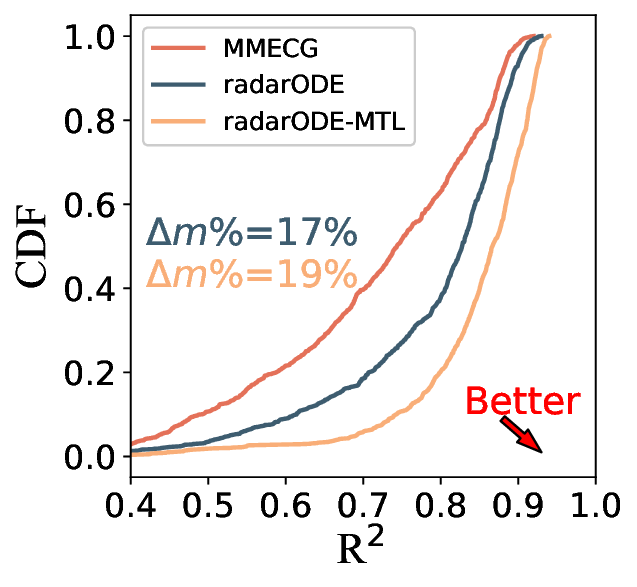}}
        \caption{Evaluations for long-term ECG recovery: (a) - (e) CDF plots of MDR, HR Error, RMSE, PCC and $R^2$, with corresponding improvements.}
        \label{fig:morph_cdfs}
\end{figure*} 

\subsubsection{Corrupt ECG Reconstruction}
The successful reconstruction in Figure~\subref*{fig:mtl_abrupt} owes to the design of radarODE-MTL with deconstructed tasks for ECG recovery. Different from other frameworks with equal length of input and output, radarODE-MTL adopts a 4-sec segment to reconstruct the ECG piece for one cardiac cycle, and the radar signal from adjacent cardiac cycles (e.g., the zoomed part in Figure~\subref*{fig:radar_abrupt}) also contributes to the recovery of the current ECG piece. In addition, if the input radar signal is fully destroyed by noise, radarODE-MTL may fail to extract any information, and the failures can be revealed by the MDR to statistically evaluate the corruptions in recovered ECG signals due to noise distortion.

The result of MDR is shown as the cumulative distribution function (CDF) in Figure~\subref*{fig:mdr_cdf} with the median MDR as $1.7\%$, $0.13\%$ and $0.13\%$ for MMECG, radarODE and radarODE-MTL respectively, and $\Delta m\%$ across $91$ trials are both $14\%$. The reason for the similar performance of two ODE-based methods is that the misaligned ECG pieces with small deviations ($<150$ms) in radarODE will not be identified as `missed detected', and hence the CDFs of MDR share a similar pattern and trend in Figure~\subref*{fig:mdr_cdf}.

\begin{figure*}[tb]
        \centering
        \subfloat[]{\label{fig:Q_cdf}\includegraphics[width=0.45\columnwidth]{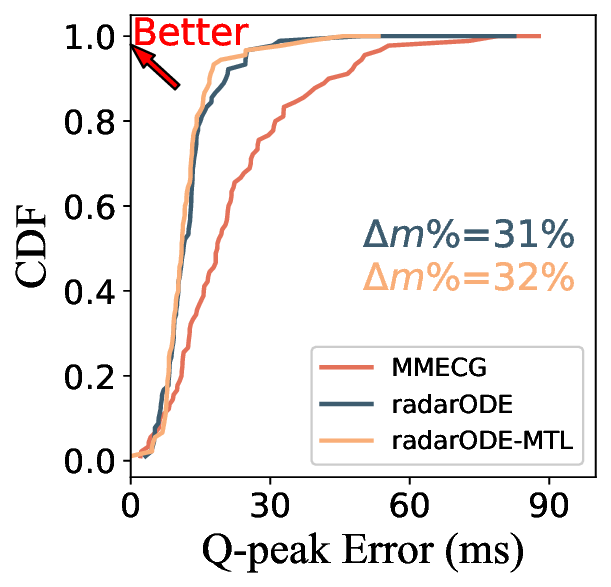}}
        \subfloat[]{\label{fig:R_cdf}\includegraphics[width=0.45\columnwidth]{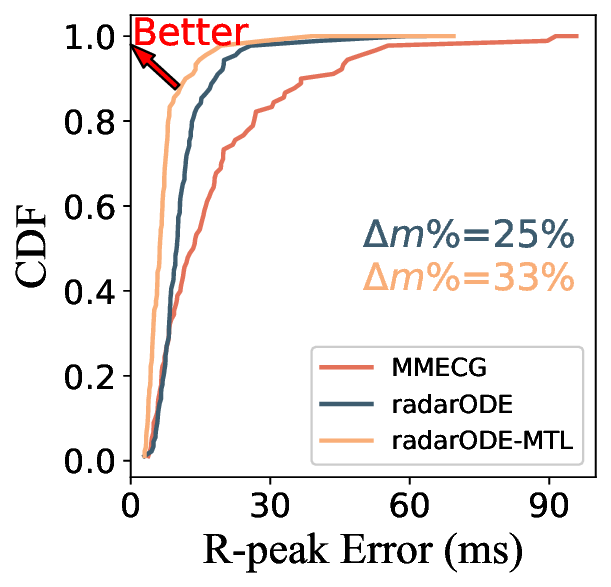}}
        \subfloat[]{\label{fig:S_cdf}\includegraphics[width=0.45\columnwidth]{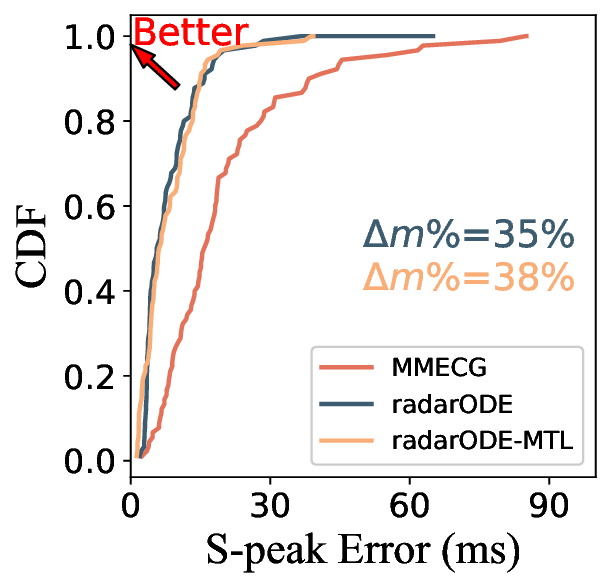}}
        \subfloat[]{\label{fig:T_cdf}\includegraphics[width=0.45\columnwidth]{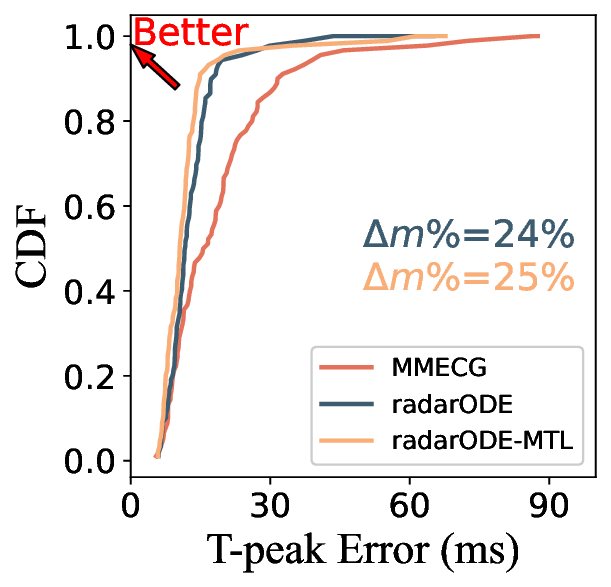}}
        \caption{{Evaluations for fine-grained ECG peaks recovery: (a) - (d) CDF plots of the timing error for QRST peaks, with corresponding improvements.}}
        \label{fig:peak_cdfs}
\end{figure*} 

\subsubsection{Coarse Cardiac Feature Reconstruction}
All three frameworks evaluated in this paper are designed for fine-grained cardiac features reconstruction and should perform well on the coarse cardiac feature (i.e., heart rate (HR) monitoring). The result in Figure~\subref*{fig:hr_cdf} coincides with the expectation with median HR error as $0.6$, $0.3$ and $0.3$ beats/min respectively, and $\Delta m\%$ for the ODE-based methods are $54\%$ and $59\%$. It is notable in Figure~\subref*{fig:hr_cdf} that the performances of ODE-based methods are very similar at the beginning, while the radarODE tends to get more errors when the noise in the raw radar signal affects the R peaks recovery, because the calculation of HR is based on the R peak positions.

\subsubsection{Fine-Grained Morphological Feature Reconstruction}
The morphological feature is an essential fine-grained feature to describe the general similarity between the recovered and ground truth ECG signals, and the morphological accuracy can be evaluated by RMSE, PCC and $R^2$, with RMSE sensitive to the peak deviation, PCC focusing on the similarity of the general shape and $R^2$ shows the interpretability of the well-trained neural network. The results are shown in Figure~\subref*{fig:rmse_cdf},~\subref*{fig:ocor_cdf} and~\subref*{fig:r2_cdf} as the CDF of RMSE/PCC/$R^2$ across $91$ trials in the dataset, and three frameworks get the median RMSE/PCC as $0.125$mV/$82.1\%$/$0.74$, $0.098$mV/$90.1\%$/$0.81$ and $0.083$mV/$92.7\%$/$0.85$ respectively.

As indicated by $\Delta m\%$, the improvements of RMSE ($28\%,33\%$) are larger than PCC ($18\%,21\%$) for radarODE and radarODE-MTL respectively, because the ODE model embedded in the decoder preserves the main features of ECG even under noises and contributes more on the peaks than on the shapes. In addition, radarODE-MTL further improves the results by aligning the ECG pieces with the predicted anchors, avoiding the misalignment issue in radarODE. Lastly, the resultant improvements in $R^2$ ($17\%$ and $19\%$ in Figure~\subref*{fig:r2_cdf}) indicate that radarODE-MTL could capture more dependency in the domain transformation of cardiac activities due to the induction of ODE model as prior knowledge, but the improvements are less than the other two metrics for morphological assessment, because $R^2$ is not sensitive to the outliers and could objectively evaluate the model ability.

\subsubsection{Fine-Grained ECG Peaks Reconstruction}
In the evaluations of timing errors of the ECG peaks it is common only to analyze QRST peaks because the inconspicuous P peaks can be miss-detected even in some ground truth signals~\cite{chen2022contactless,zhang2024radarODE}. The CDF plots for the absolute timing errors of QRST peaks are shown in Figure~\ref{fig:peak_cdfs} with the following observations:
\begin{itemize}
  \item Both ODE-based methods reveal better performance than the benchmark, but the radarODE-MTL only achieves equivalent performance as radarODE with similar $\Delta m\%$ around $31\%$, $35\%$ and $24\%$ as shown in Figure~\subref*{fig:Q_cdf},~\subref*{fig:S_cdf} and~\subref*{fig:T_cdf}. The possible reason is that radarODE-MTL only aligns the ECG pieces with R peaks, but the impacts on the QST peaks are random. In other words, the alignment of the R peak may degrade the accuracy of other peaks, and hence the overall performance of radarODE and radarODE-MTL on the QST peaks are similar.
  \item It is worth noticing that $\Delta m\%$ of the radarODE-MTL ($33\%$) on the R peak is obviously larger than that of the radarODE ($25\%$), with the median timing error as $14$, $10$ and $6$ms for three frameworks as shown in Figure~\subref*{fig:R_cdf}. Therefore, radarODE-MTL is a better way to generate long-term ECG signals by aligning the ECG pieces with predicted R peaks, instead of reintroducing the noisy time-domain radar signal as in radarODE.
\end{itemize}

\subsection{Noise Robustness Test}\label{sec:nr_test}
In this work, $10$ trials (No. $75-84$) are selected for the noise robustness test by adding different types of synthesized noises with certain decibel (dB) only in the test stage, while the training and validation stage will use the original data because adding noises into the training dataset is a data augmentation technique to improve the model performance, causing an unfair comparison in the noise robustness test~\cite{chen2024tfpred}. In addition, adding noises into the validation stage is equivalent to selecting appropriate models for the scenarios with different SNR levels and cannot prove the noise robustness of the proposed radarODE-MTL.

\subsubsection{Constant Noise}
\begin{figure*}[tb]
        \centering
        \subfloat[]{\label{fig:constant}\includegraphics[width=0.9\columnwidth]{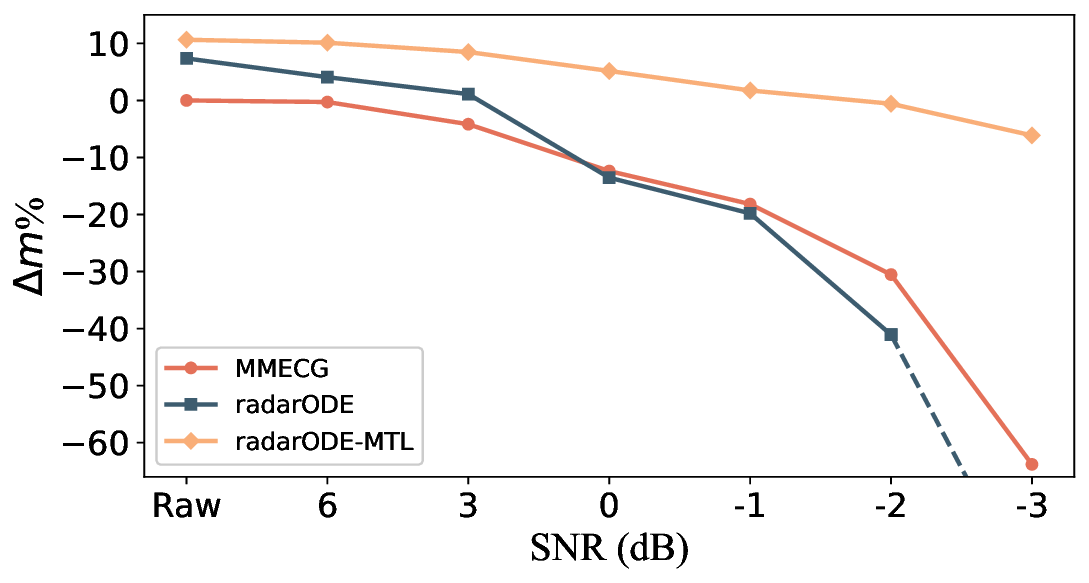}}
        \subfloat[]{\label{fig:abrupt}\includegraphics[width=0.9\columnwidth]{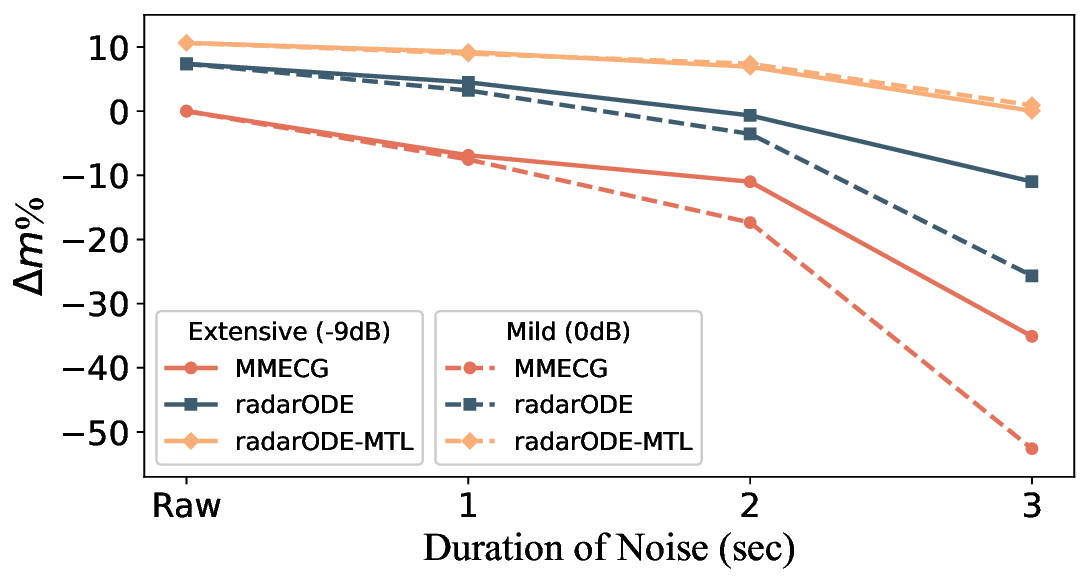}}
        \caption{Noise robustness test: (a) Impact of constant noises with different intensities, (b) Impact of abrupt noises with different intensities and durations.}
        \label{fig:noise_robu}
\end{figure*}

The constant noise normally affects the SNR of the signal and could be caused by thermal noise from electronic components or long-range detection~\cite{shen2018respiration,dong2024robust}, e.g., the SNR for the current data collection scenario will decrease to $0$dB by increasing the monitoring distance to $5$m. In the literature, low SNR scenarios can be simulated by adding Gaussian noise with different intensities as implemented in~\cite{dong2024robust,liu2024diversity,qian2023mobile,zhao2024airecg}. The baseline results for three frameworks are firstly obtained in terms of the RMSE, PCC, $R^2$, R-peak error and MDR as shown in Table~\ref{tab:mtl_compare_snr}, and $\Delta m\%$ is calculated as $0\%$, $7.47\%$ and $10.64\%$ as indicated by the initial points in Figure~\ref{fig:noise_robu}. Then, the Gaussian noises with $6$ to $ -3$dB are added into the raw radar signal without retraining the deep-learning framework, and the results are shown in Table~\ref{tab:mtl_compare_snr} with the trends of performance degradation shown in Figure~\ref{fig:noise_robu}.

\begin{table*}[t]
  \caption{Comparison of the frameworks under different SNR}
  \centering
  \begin{tabular}{lccccc|rc}
  \toprule
  \multicolumn{1}{c}{SNR} & \makecell[c]{RMSE (mV)} $\downarrow$ & PCC $\uparrow$ & $R^2$ $\uparrow$ & \makecell[c]{Peak Error (ms) } $\downarrow$ & MDR $\downarrow$ &  $\Delta m\%^1 \uparrow$ & {\makecell[c]{$P$ Value ($\times 10^{-2}$)}} \\
  \midrule
  \multicolumn{7}{c}{MMECG~\cite{chen2022contactless}} \\
  \hline
  {Baseline} & 0.107 & 83.75\% & 0.77 & 9.45 & 4.52\% & 0.00$\pm$0.55 & - \\
  $\textcolor{white}{-}6$ dB & 0.107 & 82.60\% & 0.76 & 9.76 & 4.37\% & -0.28$\pm$1.68 & 73.85 \\
  $\textcolor{white}{-}3$ dB & 0.108 & 82.64\% & 0.76 & 9.85 & 4.84\% & -4.17$\pm$1.91 & 0.04 \\
  $\textcolor{white}{-}0$ dB & 0.109 & 80.00\% & 0.74 & 11.80 & 4.92\% & -12.38$\pm$3.18 & 0.00 \\
  $-1$ dB & 0.114 & 78.55\% & 0.69 & 12.20 & 5.32\% & -18.17$\pm$4.10 & 0.00 \\
  $-2$ dB & 0.120 & 74.32\% & 0.65 & 14.64 & 5.59\% & -30.53$\pm$3.78 & 0.00 \\
  $-3$ dB & 0.127 & 62.45\% & 0.54 & 21.28 & 6.40\% & -63.81$\pm$2.15 & 0.00 \\
  \midrule
  \multicolumn{7}{c}{radarODE~\cite{zhang2024radarODE}} \\
  \hline
  {Baseline} & 0.091 & 83.53\% & 0.79& 9.08 & 4.03\% & 0.00$\pm$0.39&- \\
  $\textcolor{white}{-}6$ dB & 0.093 & 83.30\% & 0.78 & 9.12 & 4.36\% & -3.29$\pm$1.71 & 0.11 \\
  $\textcolor{white}{-}3$ dB & 0.095 & 83.01\% & 0.76 & 9.01 & 4.70\% & -6.26$\pm$1.43 & 0.00 \\
  $\textcolor{white}{-}0$ dB & 0.101 & 82.21\% & 0.69 & 9.89 & 5.86\% & -20.91$\pm$2.09 & 0.00 \\
  $-1$ dB & 0.116 & 79.66\% & 0.63& 11.90 & 5.36\% & -27.17$\pm$2.92 & 0.00 \\
  $-2$ dB & 0.157 & 70.87\% & 0.58& 13.95 & 6.19\% & -48.44$\pm$3.53 & 0.00 \\
  $-3$ dB & - & - & - & - & - & \multicolumn{1}{r}{Failed$^2$} & - \\
  \midrule
  \multicolumn{7}{c}{radarODE-MTL} \\
  \hline
  {Baseline} & 0.089 & 85.03\% & 0.81 & 8.22 & 4.08\% & 0.00$\pm$1.24 &- \\
  $\textcolor{white}{-}6$ dB & 0.088 & 85.31\% & 0.82 & 8.18 & 4.20\% & -0.52$\pm$0.72 & 19.81 \\
  $\textcolor{white}{-}3$ dB & 0.089 & 84.29\% & 0.80 & 8.31 & 4.27\% & -2.15$\pm$1.55 & 0.78 \\
  $\textcolor{white}{-}0$ dB & 0.091 & 83.77\% & 0.79 & 8.03 & 4.76\% & -5.47$\pm$2.59 & 0.05 \\
  $-1$ dB & 0.093 & 84.01\% & 0.78& 8.10 & 5.10\% & -8.89$\pm$1.21 & 0.00 \\
  $-2$ dB & 0.093 & 84.51\% & 0.79 & 8.02 & 5.45\% & -11.22$\pm$1.43 & 0.00 \\
  $-3$ dB & 0.094 & 84.96\% & 0.78& 8.19 & 6.02\% & -16.77$\pm$1.60 & 0.00 \\
  \bottomrule
  \multicolumn{8}{l}{1. $\Delta m\%$ is calculated for each framework based on each baseline.} \\
  \multicolumn{8}{l}{2. The ECG recovery fails if PCC$<60\%$, according to the empirical observation of the morphological ECG features.} \\
  \end{tabular}
  \label{tab:mtl_compare_snr}
\end{table*}

A general observation of Table~\ref{tab:mtl_compare_snr} is that all the frameworks perform well before $0$dB with a similar degradation rate as in Figure~\subref*{fig:constant}. Then, radarODE-MTL could still provide reasonable results with mild degradation after $0$dB because the MTL paradigm split the ECG reconstruction task into several sub-tasks, and each task can either be constrained by prior knowledge or leverage the information from context data with less pollution. In contrast, radarODE could generate high-fidelity ECG pieces as claimed in~\cite{zhang2024radarODE} and gets the second best baseline result in Table~\ref{tab:mtl_compare_snr}, but the design of PPI estimation stage does not consider the noise robustness. Therefore, the performance is heavily dropped to the worst in Figure~\subref*{fig:constant} because of the bad results of Peak Error as shown in Table~\ref{tab:mtl_compare_snr}. Lastly, the MMECG considers the ECG recovery as an arbitrary domain transformation problem without any constraints in the network design, and the performance also heavily degrades in Figure~\subref*{fig:constant} because only meaningless results will be generated as shown previously in Figure~\subref*{fig:mmecg_abrupt}.

\begin{table*}[t]
\centering
  \caption{Comparison of the frameworks under abrupt noises}
  \begin{tabular}{lccccc|r?ccccc|r}
  \toprule
  Duration & \makecell[c]{RMSE \\ (mV) } $\downarrow$ & PCC $\uparrow$ & $R^2$ $\uparrow$ & \makecell[c]{Peak \\ Error \\ (ms) } $\downarrow$ & MDR $\downarrow$ &  $\Delta m\%^1 \uparrow$ & \makecell[c]{RMSE \\ (mV) } $\downarrow$ & PCC $\uparrow$ & $R^2$ $\uparrow$ & \makecell[c]{Peak \\ Error \\ (ms) } $\downarrow$ & MDR $\downarrow$ &  $\Delta m\% \uparrow$\\
  \midrule
  \multicolumn{3}{l}{MMECG~\cite{chen2022contactless}:} & \multicolumn{3}{r}{Mild Body Movement ($0$ dB)} && \multicolumn{6}{c}{Extensive Body Movement ($-9$ dB)} \\
  \midrule
  {Baseline} & 0.107 & 83.75\% & 0.77 & 9.45 & 4.52\% & 0.00$\pm$0.55 & 0.107 & 83.75\% &0.77 & 9.45 & 4.52\% & 0.00$\pm$0.55 \\
  $1$ sec & 0.107 & 85.53\% & 0.78 & 10.84 & 4.82\% & -6.88$\pm$2.16 & 0.107 & 84.05\% & 0.76 & 10.93 & 4.82\% & -7.54$\pm$1.42 \\
  $2$ sec & 0.110 & 82.64\% & 0.75 & 11.31 & 5.02\% & -11.00$\pm$1.68  & 0.108 & 79.01\% & 0.68& 12.31 & 5.23\% & -17.36$\pm$2.06 \\
  $3$ sec & 0.114 & 76.87\% & 0.66 & 15.56 & 5.92\% & -35.10$\pm$2.91 & 0.116 & 75.09\% & 0.59 & 12.50 & 9.56\% & -52.61$\pm$2.15 \\
  \midrule
  \multicolumn{3}{l}{radarODE~\cite{zhang2024radarODE}:} & \multicolumn{3}{r}{Mild Body Movement ($0$ dB)} & & \multicolumn{6}{c}{Extensive Body Movement ($-9$ dB)} \\
  \midrule
  {Baseline} & 0.091 & 83.53\% & 0.79& 9.08 & 4.03\% & 0.00$\pm$0.39 & 0.091 & 83.53\% & 0.79& 9.08 & 4.03\% & 0.00$\pm$0.39 \\
  $1$ sec & 0.091 & 83.49\% & 0.79 & 9.12 & 4.36\% & -2.88$\pm$1.81 & 0.095 & 82.96\% & 0.73 & 9.15 & 4.33\% & -4.15$\pm$0.99  \\
  $2$ sec & 0.092 & 83.39\% & 0.78 & 9.82 & 4.64\% & -8.04$\pm$1.50  & 0.098 & 82.16\% & 0.70& 9.31 & 4.97\% & -10.92$\pm$1.09  \\
  $3$ sec & 0.095 & 83.01\% & 0.75& 10.01 & 5.70\% & -18.35$\pm$2.19 & 0.102 & 81.87\% & 0.68& 9.66 & 7.39\% & -33.03$\pm$2.05  \\
  \midrule
  \multicolumn{3}{l}{radarODE-MTL:} & \multicolumn{3}{r}{Mild Body Movement ($0$ dB)} && \multicolumn{6}{c}{Extensive Body Movement ($-9$ dB)} \\
  \midrule
  {Baseline} & 0.089 & 85.03\% & 0.81 & 8.22 & 4.08\% & 0.00$\pm$1.24  & 0.089 & 85.03\% & 0.81 & 8.22 & 4.08\% & 0.00$\pm$1.24  \\
  $1$ sec & 0.090 & 84.62\% & 0.80 & 7.87 & 4.42\% & -1.67$\pm$1.36 & 0.090 & 84.31\% & 0.80& 8.28 & 4.18\% &  -1.42$\pm$0.94   \\
  $2$ sec & 0.090 & 84.78\% & 0.82 & 8.29 & 4.44\% & -3.25$\pm$1.10 & 0.091 & 84.21\% & 0.79 & 8.32 & 4.41\% & -3.73$\pm$1.21  \\
  $3$ sec & 0.091 & 84.44\% & 0.78& 8.34 & 5.12\% & -9.72$\pm$1.02 & 0.095 & 84.17\% & 0.77& 8.43 & 5.10\% & -10.60$\pm$1.85  \\ 
  \bottomrule
  \multicolumn{11}{l}{1. $\Delta m\%$ is calculated for each framework based on the corresponding baseline.} \\
  \end{tabular}
  \label{tab:mtl_compare_abrupt}
\end{table*}
\subsubsection{Abrupt Noise}
In this part, the Gaussian noises with different intensities ($0$ and $-9$dB) are used to simulate mild body movement (e.g., during talking or writing) and extensive body movement (e.g., during torso movement) as suggested in the literature~\cite{chen2021movi}. In practice, the body movements have orders of magnitude larger than cardiac activities to ruin the cardiac activities, and the ability of radarODE-MTL to recover ECG signal during RBM comes from the contextual information provided by previous cardiac cycles without RBM noise. Only $20\%$ of the segments randomly selected from one trial are doped, and the duration of noise varies from $1$ to $3$ sec.

For mild body movement, the experimental results are shown in Table~\ref{tab:mtl_compare_abrupt} with the changes of $\Delta m\%$ shown in Figure~\subref*{fig:abrupt}. Firstly, it is evident that the impact of $1$-sec abrupt noise is limited for all the frameworks, and the results for ODE-based methods are almost equivalent to the baselines. Secondly, $2$-sec noise starts to have a noticeable impact on MMECG, while the ODE-based methods could preserve the performance on the morphological features (RMSE/PCC/$R^2$) with small degradation on the Peak Error and MDR. Lastly, $3$-sec noise has distorted $3/4$ of the input radar segment, and the performances of MMECG and radarODE drop obviously as shown in Figure~\subref*{fig:abrupt}, while radarODE-MTL only loses some points on $\text{MDR}=5.12\%$ as shown in Table~\ref{tab:mtl_compare_abrupt}.

In comparison, the extensive body movements with $1$ and $2$ sec have similar impacts with mild ones on ODE-based methods, because the ODE decoder could preserve the ECG shape even under strong noises, whereas the segments affected by noise cannot contribute to the recovery for MMECG as evident by the significant drop of PCC (from $84.05\%$ to $79.01\%$) as shown in Table~\ref{tab:mtl_compare_abrupt}. In addition, the $3$-sec noise destroys the ECG recovery for MMECG and radarODE with a significant degradation as shown in Figure~\subref*{fig:abrupt}, whereas the radarODE-MTL only sacrifices certain RMSE and peak accuracy with the overall degradation dropping slightly from $-9.72\%$ to $-10.60\%$ as shown in Table~\ref{tab:mtl_compare_abrupt}.

\textbf{Statistical Analysis for Noise Robustness Test: }The same T-test is also implemented for the noise robustness test as shown in Table~\ref{tab:mtl_compare_snr} and~\ref{tab:mtl_compare_abrupt}. For constant noise, $6$dB noise does not have a significant impact on the performance of MMECG and radarODE-MTL with $P>0.05$, while the performance of radarODE degrades because of the PPI error accumulation~\cite{zhang2024radarODE}. In addition, all the $P$ values for the experiments of abrupt noise are less than $0.05$ and are not listed in Table~\ref{tab:mtl_compare_abrupt}. Lastly, the impact of the noise level or duration is statistically significant, and the proposed radarODE-MTL could significantly improve the noise robustness as shown in Table~\ref{tab:mtl_compare_snr} and~\ref{tab:mtl_compare_abrupt}, because the CIs of all $\Delta m\%$ have no overlapping with MMECG or radarODE under the same noise.

In summary, the noise-robustness tests indicate that it is necessary to consider the noise robustness when designing the deep-learning model, because both MMECG and radarODE reveal a severe degradation in the performance, especially for the low SNR scenarios. In addition, the deconstruction of the ECG recovery task in radarODE-MTL could effectively resist the noises, because the ODE decoder protects the morphological feature, and the peak accuracy can be compensated from the adjacent cardiac cycles with less noise distortion.

\subsection{Complexity Analysis and Comparison}
Table~\ref{tab:complex} presents a detailed complexity comparison of three frameworks considering the parameter count (Params.), floating point operations (FLOPs), multiply-accumulate operations (MACs), and training time per epoch. The parameter count reflects the total number of parameters in each model, and FLOPs and MACs quantify the computational costs~\cite{guan2024talk2radar}. As shown in Table~\ref{tab:complex}, the complexities of ODE-based methods are higher than that of MMECG due to the different input data types, with radarODE and radarODE-MTL using spectrogram input and MMECG processing 1D radar signals. In addition, the majority of the parameters ($59\%$) and FLOPs ($95\%$) for radarODE-MTL are for the backbone stage for spectrogram processing, and an important future work is to squeeze the backbone size with reduced input spectrograms.

Compared with the gaps in model size, the training times per epoch for three frameworks are closer, because the MMECG is trained on arbitrary radar/ECG segments with a step length of $0.15$ sec~\cite{chen2022contactless}, while the ODE-based frameworks are based on single cardiac cycles. In this case, the MMECG needs to traverse $48$k samples while radarODE-MTL only has $19$k samples, indicating that many samples for MMECG training are homogeneous and cannot contribute to dataset diversity and may increase the risk of overfitting.

\begin{table}[tb]
\centering
\caption{Complexity Comparison Across Deep Learning Frameworks}
    \begin{tabular}{ll|cccc}
    \toprule
    \multicolumn{2}{c|}{Framework} & \makecell[c]{Params.\\ (M)} & \makecell[c]{FLOPs\\ (G)} & \makecell[c]{MACs\\ (G)} & \makecell[c]{Time/Epoch \\ (min)} \\
    \midrule
    \multicolumn{2}{c|}{MMECG~\cite{chen2022contactless}} & $0.67$ & $0.59$ & $0.30$ & $3.25$ \\
    \midrule
    \multicolumn{2}{c|}{radarODE~\cite{zhang2024radarODE}}& $6.04$ & $2.45$  & $1.23$  & $4.51$ \\
    \midrule
    \multirow{3}*{\makecell[c]{radarODE-\\MTL}} & Backbone & $4.81$ & $2.37$ & $1.18$ & -  \\
     &Encoder & $0.72$  & $0.05$  & $0.03$ & - \\
    &Decoder &$2.59$ & $0.07$  & $0.03$ & - \\
    &All & $8.12$ & $2.50$  & $1.23$  & $4.85$ \\
    \bottomrule
    \end{tabular}
\label{tab:complex}%
\end{table}%

\subsection{Discussions and Future work}
The proposed radarODE-MTL framework has demonstrated superior performance compared to previous approaches in generating reliable ECG signals under noisy conditions. However, potential limitations will be discussed in this subsection to motivate future enhancements in radar-based ECG recovery for practical, real-world scenarios and applications.

\subsubsection{Clock Synchronization During Data Collection}
In Figure~\ref{fig:compare_noise}, it is obvious that all the ECG R-peaks lag the radar signal peaks in the dataset~\cite{chen2022contactless}, while the actual ECG signal should lead radar signal for several milliseconds due to the electromyographic activation time (EMAT)~\cite{gao2023portable,inan2014ballistocardiography}. The misalignment is blamed for poor synchronization between the devices for collecting radar and ECG signal, while such clock synchronization is commonly neglected because the essential features of ECG (e.g., shape, peak-to-peak interval) will not be affected by EMAT~\cite{chen2022contactless,li2024radarnet,zhao2024airecg}. In future work, strict clock synchronization should be ensured during data collection to provide faithful radar-ECG pairs for the diagnosis of more diseases with irregular EMAT (e.g., heart failure syndromes and paroxysmal atrial fibrillation)~\cite{gao2023portable,inan2014ballistocardiography}.

\subsubsection{Robustness During Continuous Large-scale Body Movement}
The noise robustness test in Section~\ref{sec:nr_test} shows the better performance of radarODE-MTL compared with other frameworks, because radarODE-MTL could leverage the information from adjacent clean cardiac cycles without noise distortion. However, the recovery may still have poor quality (i.e., bad MDR in Figure~\ref{fig:abrupt}) due to the continuous large-scale body movement. For example, if the majority of the input radar signal is contaminated by strong noise without containing any clean cardiac cycle, the deep learning model may not extract any useful information for ECG recovery. In future work, advanced signal processing algorithms are necessary to be developed to ensure a high SNR signal even under continuous large-scale body movement to enable radar-based cardiac monitoring in a general scenario (e.g., walking subjects).

\section{Conclusions}\label{sec:conclusions}
This paper investigates the radar-based ECG monitoring technique and proposes a deep-learning framework radarODE-MTL to provide accurate ECG monitoring under noises. The radarODE-MTL adopts the MTL paradigm to realize the ECG reconstruction through $3$ sub-tasks, and a novel optimization strategy called EGA is also proposed to simultaneously optimize all the tasks without stall or negative transfer issues. The performance of EGA has been evaluated on various MTL tasks, and the experimental results evidence that EGA is competitive with other state-of-the-art optimization strategies on the unified task and achieves outstanding results on radar-based EGA recovery with unbalanced task difficulties. In addition, the well-trained radarODE-MTL could provide long-term ECG reconstructions with high fidelity in terms of MDR, morphological similarity and peak accuracy. Lastly, this is the first study that conducts noise-robustness tests for deep-learning frameworks, and the proposed radarODE-MTL could also achieve reasonable ECG recovery with mild degradation under constant and abrupt noises. In the future, the recovery of P peaks in ECG should be considered for the potential diagnosis of cardiovascular diseases with abnormal ECG waveforms (e.g., atrial fibrillation and heart block), and transfer learning or data augmentation might be adopted for alleviating data scarcity for patients.

\end{document}